\documentclass[3p,times,twocolumn]{elsarticle}

\usepackage{ecrc-ac}

\volume{00}
\firstpage{1}
\journalname{Nuclear Physics B Proceedings Supplement}
\runauth{M.\ Kr\"amer and M.\ M\"uhlleitner}

\jid{nuphbp}
\jnltitlelogo{Nuclear Physics B Proceedings Supplement}

\usepackage{amssymb}
\usepackage[figuresright]{rotating}

\usepackage{amsmath}

\newcommand{\beq}{\begin{equation}}
\newcommand{\eeq}{\end{equation}}
\newcommand{\bea}{\begin{eqnarray}}
\newcommand{\eea}{\end{eqnarray}}
\newcommand{\bed}{\begin{displaymath}}
\newcommand{\eed}{\end{displaymath}}

\newcommand{\tga}{\tan\alpha}
\newcommand{\tgb}{\tan\beta}
\newcommand{\stgb}{{\rm tg}^2\beta}
\newcommand{\sgl}{\tilde{g}}

\newcommand{\ssb}{\tilde b}

\usepackage{multirow}
\newcommand{\lsim}{\raisebox{-0.13cm}{~\shortstack{$<$ \\[-0.07cm] $\sim$}}~}

\biboptions{compress}

\begin{document}

\begin{frontmatter}



\dochead{\small{\flushleft{\vspace*{-25mm}KA-TP-01-2015\\SFB/CPP-14-115\\SLAC-PUB-16201\\TTK-15-05\\[5mm]}}}

\title{Higgs Physics}


 \author[rwth,slac]{Michael Kr\"amer} \author[kit]{Margarete M\"uhlleitner} 
 
\address[rwth]{Institute for Theoretical Particle Physics and Cosmology, RWTH Aachen University, D-52056 Aachen, Germany}
\address[slac]{SLAC National Accelerator Laboratory, Stanford University, Stanford, CA 94025, USA}
\address[kit]{Institute for Theoretical Physics, Karlsruhe Institute of Technology, 76128 Karlsruhe, Germany}

\begin{abstract}
  We discuss the interpretation of the LHC
  Higgs data and the test of the Higgs mechanism. This is
  done in a more model-independent approach relying on an effective
  Lagrangian, as well as in specific models like composite
  Higgs models and supersymmetric extensions of the Standard Model. The proper 
  interpretation of the data requires the inclusion of higher-order
  corrections both for the relevant Higgs parameters and the
  production and decay processes. We review recent results 
obtained within the Collaborative Research Centre / Transregio 9 ``Computational Particle Physics''.
\end{abstract}

\begin{keyword}
Higgs physics, beyond the Standard Model 
\end{keyword}

\end{frontmatter}


\section{Introduction}
\label{sec:intro}
With the announcement of the discovery of a new scalar particle by the
LHC experiments ATLAS \cite{:2012gk} and CMS
\cite{:2012gu} particle physics has entered a new 
era. The discovery immediately triggered activities to study the
nature of this particle. The determination of its properties like the 
spin and parity quantum numbers and the couplings to other Standard
Model (SM) particles strongly suggest it to be a Higgs boson, {\it
  i.e.}~the particle responsible for electroweak symmetry breaking
(EWSB). From the available data, however, it cannot be concluded yet
that it is the SM Higgs boson. There is still room for interpretations
within numerous extensions beyond the SM (BSM). At the same time, there 
is no hint for the existence of new particles that might shed light on
the true nature of EWSB, which can be weakly or strongly
interacting. In view of no direct observation of new states, an
effective Lagrangian for the light boson is needed that parametrizes
our ignorance of the EWSB sector. Such an effective description is
valid as long as New Physics (NP) states appear at a scale much larger
than the Higgs boson mass, $M \gg m_h$. 

At the LHC the Higgs boson couplings cannot be measured without
applying model-assumptions. Only ratios of branching ratios,
respectively, of couplings are accessible. In order to extract 
information on the Higgs boson couplings, fits are performed to the
experimentally measured values of the signal strengths $\mu$. 
The signal strengths $\mu$ quantify the Higgs signal rates in a specific final state $X$ with respect to the corresponding SM
expectation, and hence are equal to one for the SM Higgs boson. The
effective Lagrangian provides a tool that allows to consistently
depart from the SM and to calculate BSM rates that can then be used in
the fit to the $\mu$ values. 

The ultimate step in the experimental verification of the Higgs sector is
the measurement of the Higgs self-couplings. With the 
experimentally extracted self-couplings the Higgs potential can be
reconstructed and its typical shape with a non-vanishing vacuum
expectation value (VEV) can be tested. While the trilinear Higgs
self-coupling is accessible in Higgs pair production, the extraction
of the quartic self-interaction from triple Higgs production is beyond
the scope of existing and future collider experiments. 
Due to the smallness of the double Higgs production cross sections and
because of large backgrounds 
the extraction of the trilinear Higgs self-coupling will be
challenging and requires highest-possible energies and luminosities. 

In this report we will review Higgs boson phenomenology at the
LHC. We first investigate in Section \ref{sec:higgsfits} 
a general parametrization of BSM Higgs
physics in terms of an effective Lagrangian approach and 
its implementation into automatic tools for the
calculation of Higgs decay rates and production cross
sections. Furthermore, its application in fits to
the experimentally 
measured $\mu$ values will briefly be discussed. Section
\ref{sec:higgsself} is dedicated to the dominant Higgs pair production
processes at the LHC and the higher-order corrections that we have
calculated in order to improve the theoretical predictions. We give
error estimates for the various processes and present the outcome of a
parton level analysis in the gluon fusion process. The next two
sections are devoted to Higgs interpretations in BSM models based
on strong and weak interaction dynamics. In Section
\ref{sec:composite} we discuss the phenomenology of composite Higgs
models as an example for strong EWSB. Section \ref{sec:susy} is
devoted to supersymmetric extensions of the Higgs sector, \textit{i.e.}~the
minimal (MSSM) and Next-to-Minimal Supersymmetric extension of the SM (NMSSM). The last
section \ref{sec:couplings} closes the review by investigating Higgs
coupling measurements at present and future colliders and what can be
learned from these with respect to NP extensions beyond the SM. 
%
\section{Interpretation of LHC Higgs Data}
\label{sec:higgsfits}
The effective Lagrangian for the description of NP effects that emerge
at scales far beyond the EWSB scale is based on an expansion in the
number of fields and derivatives. Its detailed form depends on the
assumptions that are applied. In view of the present LHC data it is
reasonable to assume the Higgs to be CP-even and part of an $SU(2)_L$
doublet as well as baryon- and lepton number conservation. The leading
NP effects are then given by 53 operators with dimension 6
\cite{Burges:1983zg,Leung:1984ni,Buchmuller:1985jz,Grzadkowski:2010es},
considering only one family of quarks and leptons. In case of
CP-violation there are another 6 operators. In the
strongly-interacting light Higgs (SILH) basis the effective Lagrangian
composed of the SM Lagrangian ${\cal L}_{\rm SM}$ and the
contributions from dimension-6 operators $O_i$ is given by \cite{Giudice:2007fh, Grinstein:2007iv}
\begin{eqnarray}
{\cal L} &=& {\cal L}_{\rm SM}  + \sum_i \bar c_i O_i 
\equiv {\cal L}_{\rm SM} + \Delta {\cal L}_{\rm SILH}  \nonumber \\
&+& \Delta {\cal L}_{F_1}
+  \Delta {\cal L}_{F_2}  + \Delta {\cal L}_{V} + \Delta {\cal L}_{4F}\, .
\label{eq:effL}
\end{eqnarray}
The explicit form of ${\cal L}_{\text{SM}}$ and the Lagrangians 
$\Delta {\cal L}_i$ containing the dimension-6 operators, along with
the conventions for the covariant derivatives and the gauge field
strengths can be found in Ref.~\cite{Contino:2013kra}. The Wilson
coefficients $\bar{c}_i$ are matrices in flavor space, and a
summation over flavor indices has been implicitly assumed. Each of the 
operators $O_{u,d,l}$ is assumed  to be flavor-aligned with the
corresponding mass term in order to avoid Flavor-Changing Neutral
Currents (FCNC) through the tree-level exchange of the Higgs
boson. This leads to the coefficients $\bar{c}_{u,d,l}$ being
proportional to the identity matrix in flavor space. Assuming also
CP-invariance they are taken to be real. Applying the power counting
of Ref.~\cite{Giudice:2007fh}, a naive estimate of the Wilson
coefficients $\bar{c}_i$ has been given in \cite{Contino:2013kra}. In
the unitary gauge with canonically normalized fields, the SILH
effective Lagrangian $\Delta {\cal L}_{\rm SILH}$ can be cast into the form
\begin{eqnarray} 
{\cal L} &=&
\frac{1}{2} \partial_\mu h\ \partial^\mu h - \frac{1}{2} m_h^2 h^2 
- c_3 \, \frac{1}{6} \left(\frac{3 m_h^2}{v}\right) h^3 \nonumber \\
&-& \sum_{\psi = u,d,l} m_{\psi^{(i)}} \, \bar\psi^{(i)}\psi^{(i)}
\left( 1 + c_\psi \frac{h}{v} + \dots \right) \nonumber
\\[0.cm]
&+& m_W^2\,  W^+_\mu W^{-\, \mu} \left(1 + 2 c_W\, \frac{h}{v} + \dots
\right) \nonumber \\
&+& \frac{1}{2} m_Z^2\,  Z_\mu Z^\mu \left(1 + 2 c_Z\, \frac{h}{v} +
  \dots \right) + \dots \nonumber
\\[0.3cm]
&+& \left(  c_{WW}\,   W_{\mu\nu}^+ W^{-\mu\nu}  + \frac{c_{ZZ}}{2} \,
  Z_{\mu\nu}Z^{\mu\nu} +  c_{Z\gamma} \, Z_{\mu\nu} \gamma^{\mu\nu} \right.
  \nonumber \\
&&\left. + \frac{c_{\gamma\gamma}}{2}\, \gamma_{\mu\nu}\gamma^{\mu\nu}
  + \frac{c_{gg}}{2}\, G_{\mu\nu}^aG^{a\mu\nu} \right) \frac{h}{v} \nonumber
\\[0.2cm]
&+&  \Big( \left(c_{W\partial W}\, W^-_\nu D_\mu W^{+\mu\nu}+
  h.c.\right)+c_{Z\partial Z}\,  Z_\nu\partial_\mu Z^{\mu\nu}
\nonumber \\
&& +  c_{Z\partial \gamma}\, Z_\nu\partial_\mu\gamma^{\mu\nu} \Big)\,
\frac{h}{v} + \dots
\label{eq:chiralL}
\end{eqnarray}
Shown are terms with up to three fields and at least one Higgs
boson. The couplings $c_i$ can be expressed as functions linear in the
Wilson coefficients of the effective Lagrangian
Eq.~(\ref{eq:effL}). The explicit relations can be found in Table~1 of
Ref.~\cite{Contino:2014aaa}. In particular, as a consequence of the accidental
custodial invariance of the SILH Lagrangian at the dimension-6 level
\cite{Contino:2013kra} the following two relations hold
\begin{eqnarray}
\label{eq:identity1}
c_{WW}  - c_{ZZ} \cos^2\!\theta_W &=& c_{Z\gamma} \sin 2\theta_W +
c_{\gamma\gamma}  \sin^2\!\theta_W \nonumber \\[0.4cm] 
\label{eq:identity2}
c_{W\partial W}  - c_{Z\partial Z} \cos^2\!\theta_W &=&
\frac{c_{Z\partial \gamma}}{2} \sin 2\theta_W  \, .  
\end{eqnarray}
A third relation, $c_W = c_Z $, 
holds if additionally custodial invariance is required for $\Delta
{\cal L}_{\rm SILH}$. Allowing for arbitrary values of the couplings
$c_i$, Eq.~(\ref{eq:chiralL}) is the most general effective Lagrangian at
${\cal O}(p^4)$ in a derivative expansion by focusing on cubic terms
with at least one Higgs boson and making two further
assumptions, which are CP conservation and vector fields that couple
to conserved currents. The Higgs field in Eq.~(\ref{eq:chiralL}) needs
not be part of an electroweak doublet, so that the Lagrangian can be
considered as a generalization of $\Delta {\cal L}_{\rm SILH}$. It
contains 10 couplings involving a single Higgs boson and two gauge
fields ($hVV$ couplings, with $V = W,Z,\gamma,g$), 
3 linear combinations of which  vanish if custodial symmetry is
imposed~\cite{Contino:2013kra}. If the electroweak (EW) symmetry is realized
non-linearly, all Higgs couplings are independent of other parameters
not involving the Higgs boson. In a linear realization, however, only
4 $hVV$ couplings are independent of the other EW measurements
\cite{Elias-Miro:2013mua}.  Therefore in custodial invariant scenarios
it is impossible, by focusing only on $hVV$ couplings, to tell whether
the Higgs boson is part of an EW doublet. The doublet nature may be
disproved by the decorrelation between the $hVV$ couplings and the
other EW data \cite{Isidori:2013cga, Brivio:2013pma,Pomarol:2013zra,Gupta:2014rxa}. 

The parametrizations of BSM couplings in terms of the dimension-6
extension given by $\Delta {\cal L}_{\rm SILH}$ as well as the one given by the
non-linear Lagrangian Eq.~(\ref{eq:chiralL}) have been implemented in the
Fortran code \textsc{Ehdecay}\footnote{The program can be
downloaded at the URL:\newline
http://www-itp.particle.uni-karlsruhe.de/~maggie/eHDECAY/.}
\cite{Contino:2014aaa}. Furthermore two
benchmark composite Higgs models MCHM$_4$~\cite{Agashe:2004rs} and
MCHM$_5$~\cite{Contino:2006qr} have been included. It is based on the
code \textsc{Hdecay} \cite{Djouadi:1997yw,Djouadi:2006bz} for the
computation of decay widths and branching ratios in the SM and the
minimal supersymmetric extension of the SM. All relevant QCD
corrections, which generally factorize with respect to the expansion
in the number of fields and derivatives of the effective Lagrangian,
have been included by using the existing SM computations. The
inclusion of the electroweak corrections in a consistent way is only possible in the
framework of the SILH Lagrangian and up to higher orders in $(v/f)$. Here $v$ is
the weak scale $v \approx 246$~GeV, and $f\equiv
M/g_\star$ is given in terms of the NP scale $M$ and
the typical coupling $g_\star$ of the NP sector. In \textsc{Ehdecay} the user can choose to take the EW corrections into
account in case of the SILH parametrization. 

With the effective Lagrangian Eq.~(\ref{eq:chiralL}) at hand, the $\mu$ rates can be
calculated by smoothly departing from the SM in a consistent way. The
approach followed here at present
\cite{LHCHiggsCrossSectionWorkingGroup:2012nn} is to assume that, while the
couplings are modified by an overall modification factor, the coupling
structures are not modified with respect to the SM.\footnote{The impact of higher-dimensional operators 
on differential distributions, and the limitations of an effective Lagrangian approach have recently been discussed in Ref.~\cite{Biekoetter:2014jwa}.}
 Furthermore, the
narrow width approximation is applied, factorizing the production and
decay processes. Finally, the
Higgs signal observed at the LHC is supposed to be built up by a single
resonance. Our fits performed to the experimentally measured $\mu$
values \cite{Espinosa:2012ir,Espinosa:2012im} show that the SM Higgs
boson is compatible with the LHC Higgs data within 2$\sigma$. Our fits to
a possible invisible branching ratio
\cite{Espinosa:2012im,Espinosa:2012vu} reveal that large invisible
branching ratios are consistent with the LHC measurements. 
%
\section{Higgs Pair Production at the LHC}
\label{sec:higgsself}
The measurements of the trilinear and quartic Higgs self-couplings
allow for the reconstruction of the Higgs potential
\cite{Djouadi:1999gv,Djouadi:1999rca} and thereby the
experimental verification of its typical form with a non-vanishing
VEV, required for the Higgs mechanism to work. Higgs pair production
gives access to the trilinear Higgs self-coupling. The dominant
processes for double Higgs production at the LHC are: 
\begin{itemize}
\item[$(i)$] the gluon fusion mechanism $gg\to HH$
\cite{Eboli:1987dy,Glover:1987nx,Dicus:1988yh,Plehn:1996wb};
\item[$(ii)$] the $WW/ZZ$ fusion processes (VBF), $qq' \to V^* V^* qq' \to HH
qq'$ ($V=W,Z$)
\cite{Eboli:1987dy,Keung:1987nw,Dicus:1987ez,Dobrovolskaya:1990kx,Abbasabadi:1988ja}; 
\item[$(iii)$] the double Higgs-strahlung process, $q\bar{q}' \to V^* \to V
HH$
($V=W,Z$) \cite{Barger:1988jk}; 
\item[$(iv)$] associated production of two Higgs bosons with a top quark
pair, $pp \to t\bar{t} HH$ \cite{Moretti:2004wa}.
\end{itemize}
The loop-induced gluon fusion process provides the dominant
contribution to Higgs pair production. It has been calculated at
next-to-leading order (NLO) QCD in an effective field theory (EFT)
approximation \cite{Plehn:1996wb} by applying the low-energy theorem
\cite{Ellis:1975ap,Shifman:1979eb,Kniehl:1995tn}, {\it i.e.}~by
assuming infinitely heavy quarks. The NLO cross section for a 125~GeV
Higgs boson amounts to $33.89$~fb at a center-of-mass (c.m.) energy
of 14~TeV. In obtaining this number the leading order (LO) cross
section has been calculated including the full mass dependence in
order to improve the perturbative results. The NLO $K$-factor, {\it
  i.e.}~the ratio between the NLO and the LO results is large and
amounts to 1.5-2 depending on the c.m.~energy. The error estimate
based on the missing higher order corrections, PDF$+\alpha_s$
uncertainties and the application of the EFT approach results in
an error of ${\cal O}(30\%)$ at 14~TeV \cite{Baglio:2012np}. 
In the meantime top-quark mass effects at NLO have been estimated
\cite{Grigo:2013rya,Grigo:2013xya,Maltoni:2014eza} and results at
next-to-next-to-leading-order (NNLO) in the heavy top mass
approximation \cite{deFlorian:2013uza,deFlorian:2013jea,Grigo:2014jma} are available. 

The next important di-Higgs production process is vector boson
fusion into Higgs pairs, closely followed by Higgs pair production in
association with a top-quark pair, which is known at LO QCD only. For
the VBF process we have calculated the NLO QCD corrections
\cite{Baglio:2012np} in the structure function approach in complete
analogy to the single Higgs VFB process \cite{Figy:2003nv}. Setting
the renormalization and factorization scale equal to the momentum of
the exchanged weak boson we found an increase of $\sim +7$\% of the
total cross section with respect to the LO result. For the smallest
di-Higgs production process through Higgs-strahlung we have updated
the NLO results in analogy to single Higgs-strahlung
\cite{Altarelli:1979ub,Han:1991ia,KubarAndre:1978uy} and computed the
QCD corrections at NNLO \cite{Baglio:2012np}. At NNLO there is a
new gluon-initiated contribution to $ZHH$ production to be taken into
account. It turns out that in contrast to single Higgs production
\cite{Dicus:1988yh,Brein:2003wg,Kniehl:1990iva}, its contribution at
NNLO is sizeable. 

\begin{figure}[t]
\begin{center}
\includegraphics[scale=0.6]{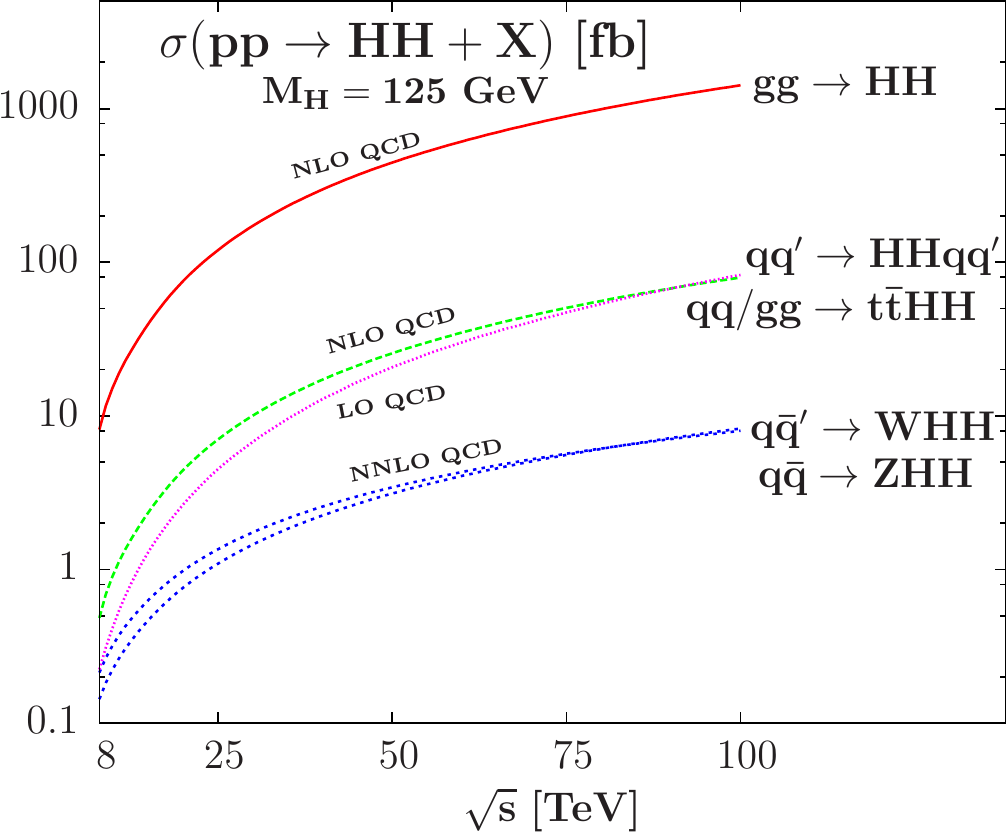}
\end{center}
\it{\vspace{-6mm}\caption{The total cross sections for Higgs pair
    production at the LHC, including
    higher order corrections, in the main channels -- gluon fusion (red/full), VBF
    (green/dashed), Higgs-strahlung (blue/dotted), associated
    production with $t\bar t$ (violet/dotted with small dots)
    -- as a function of the c.m. energy with $M_H=125$
    GeV. The MSTW2008 PDF set has been used and higher order
    corrections are included. Figure taken from \cite{Baglio:2012np}
    where also details can be found. \label{fig:hpairprodcxn}}}
\end{figure}

Figure~\ref{fig:hpairprodcxn} shows the total LHC cross sections for
the four classes of Higgs pair production as a function of the
c.m.~energy. The central renormalization and factorization scales,
$\mu_R$, respectively, $\mu_F$, that have been used are ($\mu_R =
\mu_F = \mu_0$) 
\begin{eqnarray}
\hspace*{-5mm}\mu_0^{gg\to HH} &=& M_{HH},\,\,\, \mu_0^{qq'\to HHqq'} =
Q_{V^*},\,\,\, \\
\hspace*{-5mm}\mu_0^{q\bar{q}'\to VHH} &=& M_{VHH},\,\,\,
\mu_0^{q\bar{q}/gg\to t\bar t HH} = M_t\!+\! \frac12 M_{HH}, \nonumber
\end{eqnarray}
where $M_{HH}$ denotes the invariant mass of the Higgs pair.
Note that all pair production cross sections are $\sim 1000$ times
smaller than the corresponding single Higgs production channels. High
luminosities are therefore required to probe the Higgs pair production
channels at the LHC. 
The smallness of the cross sections together with the large QCD
backgrounds at the LHC make the analysis of di-Higgs production
challenging. We have performed a parton-level analysis for the
dominant gluon fusion process into Higgs pairs in different final
states, which are $b\bar{b}\gamma\gamma$, $b\bar{b}\tau\bar{\tau}$ and
$b \bar{b} W^+ W^-$ with the $W$ bosons decaying leptonically. While the
$b \bar{b} W^+ W^-$ final state turns out not to be promising after
applying acceptance and selection cuts, the significances in the
$b\bar{b}\gamma\gamma$ and $b\bar{b}\tau\bar{\tau}$ final states reach
$\sim 16$ and $\sim 9$, respectively, after applying the cuts. The
event numbers are not too small so that they are promising enough to
start a real experimental analysis taking into account detector and
hadronization effects. The sensitivity of the various pair production
processes to the trilinear Higgs coupling is diluted by additional
diagrams to the processes that do not involve the Higgs
self-interaction. The sensitivity can be studied by varying the
self-coupling in terms of the SM trilinear coupling by a scale factor
$\kappa$. The most sensitive channel turns out to be by far the VBF
production mode. Taking into account theoretical and statistical
uncertainties in the pair production channels the trilinear Higgs
self-coupling may be expected to be measured within a factor of two. 

\section{Composite Higgs Models}
\label{sec:composite}

\begin{figure*}[t!]
\begin{center}
\includegraphics[width=0.25\linewidth,angle=-90]{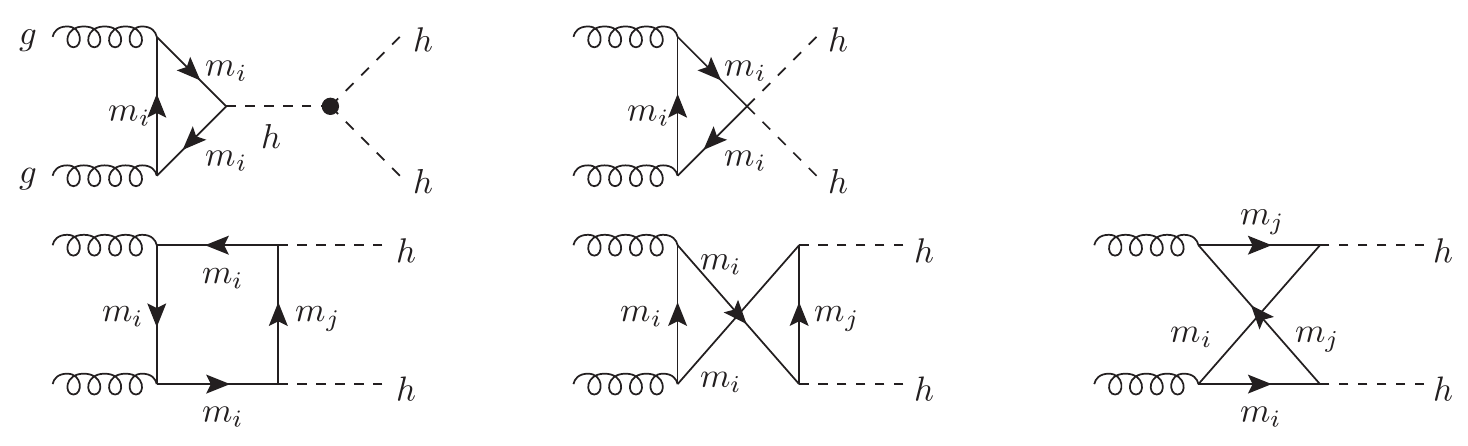}
\caption{Generic diagrams contributing to double Higgs
  production via gluon fusion in composite Higgs models with $n_f$ novel fermionic resonances of mass $m_{i}$ ($i=1,...,n_f$). The index $j$ is introduced to
indicate that different fermions can contribute to each box
diagram. From~Ref.~\cite{Gillioz:2012se}.}
\label{fig:2higgscomp}
\end{center}
\vspace*{-0.5cm}
\end{figure*}
Composite Higgs Models are examples of models with a strong dynamics
underlying EWSB. In these models a light Higgs boson arises as a pseudo
Nambu-Goldstone boson from a strongly-interacting sector
\cite{Kaplan:1983fs,Dimopoulos:1981xc,Banks:1984gj,Kaplan:1983sm,Georgi:1984ef,Georgi:1984af,Dugan:1984hq}. This
implies modified Higgs couplings with respect to the SM. In
Ref.~\cite{Giudice:2007fh} an effective low-energy description of a
Strongly Interacting Light Higgs boson has been given, which can be
viewed as the first term of an expansion in $\xi = (v/f)^2$, with the
EWSB scale $v$ and the scale of the strong dynamics $f$. This SILH
Lagrangian can be used in the vicinity of the SM limit given by $\xi
\to 0$. Larger values of $\xi$ as {\it e.g.} the technicolor limit,
$\xi \to 1$, require a resummation of the series in $\xi$. This is provided
by explicit models built in five-dimensional warped space. In the
Minimal Composite Higgs Models (MCHM) of Refs.~\cite{Agashe:2004rs,
  Contino:2006qr} the global symmetry $SO(5)\times U(1)$ is broken
down at the scale $f$ to $SO(4)\times U(1)$ on the infrared brane and
to the SM $SU(2)_L\times U(1)_Y$ on the ultraviolet brane. The
modifications of the Higgs couplings in these models can then be
described by one single  parameter, given by $\xi$. The modification
factor of the Higgs coupling to fermions depends on the representations of the bulk
symmetry into which the fermions are embedded. In the model of
Ref.~\cite{Agashe:2004rs}, called $\text{MCHM}_4$, the fermions are in
the spinorial representation of $SO(5)$, in the model $\text{MCHM}_5$ of
Ref.~\cite{Contino:2006qr} they are in the fundamental
representation. The question of the generation of fermion masses in
composite Higgs models is solved by the hypothesis of partial
compositeness \cite{Kaplan:1991dc,Contino:2006nn}. It assumes that the
SM fermions, which are elementary, couple linearly to heavy states of
the strong sector with the same quantum numbers, implying in
particular the top quark to be largely composite. These couplings
explicitly break the global symmetry of the strong sector. The Higgs
potential is generated from loops of SM particles with EWSB triggered
by the top loops which provide the dominant contribution. The Higgs
self-couplings therefore also depend on the representation of the
fermions, and the Higgs boson mass is related to the fermion
sector. It has been shown that a low-mass Higgs boson of $\sim 125$
GeV can naturally be accommodated only if the heavy quark partners are
rather light, {\it i.e.} for masses below about 1 TeV, see
\cite{Contino:2006qr,Matsedonskyi:2012ym,Redi:2012ha,Panico:2012uw,Pappadopulo:2013vca,Marzocca:2012zn,Pomarol:2012qf}. 
Phenomenologically the modified Higgs couplings to the SM particles
change the Higgs production and decay rates
\cite{Contino:2014aaa,Espinosa:2010vn}. Another important consequence
is the increase of the cross section for double Higgs production in
vector boson fusion with the energy
\cite{Giudice:2007fh,Contino:2010mh,Grober:2010yv,Contino:2013gna}. Finally
due to the coupling modifications, composite Higgs models are challenged by
electroweak precision tests (EWPTs)
\cite{Giudice:2007fh,Agashe:2005dk,Barbieri:2007bh,Pomarol:2008bh,Contino:2010rs}. The
tension with the $S$ and $T$ parameters \cite{Peskin:1991sw} can be
weakened through the contributions from new heavy fermions
\cite{Lavoura:1992np,Lodone:2008yy,Gillioz:2008hs,Anastasiou:2009rv,Grojean:2013qca,Gillioz:2013pba}. 

So far we do not have any direct evidence for new BSM
particles. Indirect information on heavy fermion partners could in
principle also be obtained from Higgs physics. These new states would
contribute to the loop-induced couplings of the Higgs boson to gluons
and photons. It is also these couplings that are involved in the most
sensitive Higgs search channels. The structure of these couplings can
effectively be described by the Low-Energy Theorem (LET). In
\cite{Gillioz:2012se} we have extended the 
theorem to take into account possible non-linear Higgs interactions as
well as new states resulting from a strong dynamics at the origin of
the EWSB. The dominant Higgs production process at the LHC, gluon
fusion, is mediated by a top loop with a subleading contribution from
the bottom loop. In composite Higgs models with heavy colored fermions
and sizable couplings to the Higgs boson, these new states give additional
loop contributions that should be taken into account. It has been
shown that in explicit constructions gluon fusion into a single Higgs
boson, $gg\to h$, computed in the LET approximation, is insensitive to
the details of the heavy fermion spectrum
\cite{Falkowski:2007hz,Low:2010mr,Azatov:2011qy,Furlan:2011uq}. The 
cross section only depends on the ratio $v/f$, but not on the couplings and the
masses of the composite fermions. Although the top Yukawa
coupling receives a correction due to the mixing with heavy resonances, which
depends on the composite couplings, this contribution is exactly
canceled by the loops of the extra fermions, so that the cross section
only depends on $v/f$. This holds both in models with partial
compositeness and in Little Higgs theories. The gluon fusion cross
section in the composite Higgs model can therefore be obtained from
the SM result by simply multiplying the latter with the squared
rescaling factor of the Yukawa couplings, that comes from the
non-linearity of the model. Every correction due to the fermionic
resonances can be neglected. This insensitivity to the composite
couplings, however, holds exactly only in the LET approximation. There
are corrections due to finite fermion mass effects. The LET
approximates the gluon fusion Higgs production rate very well, and it
indeed turns out that these corrections are very small, even for large
values of the compositeness parameter $\xi$. 

Another process mediated by colored fermion loops is double Higgs
production through gluon fusion. In composite Higgs models the process
is changed in two 
ways compared to the SM. First, there is a genuinely new contribution to the
amplitude from a new coupling between two Higgs bosons and two
fermions, which arises from the non-linearity of the strong sector. In
the SM limit this $f\bar{f}hh$ coupling vanishes. Second, the effects
from the top partners in the loop have to be taken into account. They
also give rise to a new box diagram that involves off-diagonal Yukawa 
couplings. These Yukawa couplings only involve the top quark and its
charge 2/3 heavy composite partners. The diagrams contributing to the
process are
shown in Fig.~\ref{fig:2higgscomp}. In Ref.~\cite{Grober:2010yv} a first study on
Higgs pair production in composite Higgs models was performed,
neglecting top partners. It was found that the cross section can be
significantly enhanced due to the new $t\bar{t}hh$ coupling. 
An explicit calculation of the di-Higgs production cross section in
the LET approximation shows that it is insensitive to composite
couplings. The reason is a cancellation completely analogous to the
single Higgs production case. 
The comparison with
Ref.~\cite{Grober:2010yv}, where the full top mass dependence was
retained, shows that the LET underestimates the true result
considerably, however. 

In the SM it has been known that the $m_t \to \infty$
limit approximates the full result only within 20\% \cite{Plehn:1996wb}. Moreover
it gives rise to incorrect kinematic distributions \cite{Baur:2002rb}. The same
is observed in composite Higgs models, and in particular in MCHM$_5$
the validity of the expansion gets even worse.
This behaviour can be understood by looking at the expansion
parameter. In single Higgs production this is $m_h^2/(4m_t^2)$ and the
series converges quickly. In double Higgs production, however, the
expansion is performed in $\hat{s}/(4m_t^2)$ with the partonic
c.m.~energy squared $\hat{s} \gg 4m_h^2$ which is not small, so that
the expansion is not as good as in the single Higgs case. In MCHM$_5$
the presence of the new triangle diagram containing the $t\bar{t}hh$
coupling, which contrary to the triangle diagram with the virtual
Higgs boson exchange does not vanish for large $\hat{s}$, renders the
convergence of the expansion even worse. 

This shall be exemplified for
an explicit composite Higgs model, MCHM$_5$, with extra fermionic
resonances. It is based on the symmetry pattern $SO(5)\times
U(1)_X/SO(4) \times U(1)_X$. The additional local symmetry $U(1)_X$ is
introduced in order to correctly reproduce the fermion charges. The SM
electroweak group $SU(2)_L \times U(1)_Y$ is embedded into
$SO(4)\times U(1)_X$ and the hypercharge $Y$ is then given by
$Y=T_{R}^{3}+X$ \cite{Agashe:2004rs,Contino:2006qr}. 
The vector-like fermions introduced in the model have quantum numbers
such that they can mix linearly with the SM fermions, the left-handed
doublet and right-handed singlet of the third generation $q_L = (t_L,
b_L)^T$ and $t_R$.  At the same time they have 'proto-Yukawa' interactions with
the composite Higgs. The composite fermions are required to transform
as a complete ${\bf 5}_{2/3}$ under $SO(5) \times U(1)_X$ . In this
representation no dangerously large tree-level corrections to the
$Zb_L b_L$ coupling arise, provided a discrete symmetry $P_{LR}$ exchanging the
$SU(2)_L$ and $SU(2)_R$ factors is imposed~\cite{Contino:2006qr}.
Under $SU(2)_L \times SU(2)_R$  a $\mathbf{5}$ of $SO(5)$
decomposes as $\mathbf{5}\sim (\mathbf{2},\mathbf{2})\oplus
(\mathbf{1},\mathbf{1})\,$. The bi-doublet $(\mathbf{2},\mathbf{2})$ is
formed by the $SU(2)_{L}$ doublets $Q=(T,B)^{T}$ and
$X=(X^{5/3},X^{2/3})^{T}$, while $\tilde{T}$ is a singlet
$(\mathbf{1},\mathbf{1})\,$ under $SU(2)_L\times SU(2)_R$. 

\begin{figure*}[t]
	\centering
	\includegraphics[width=0.45\linewidth]{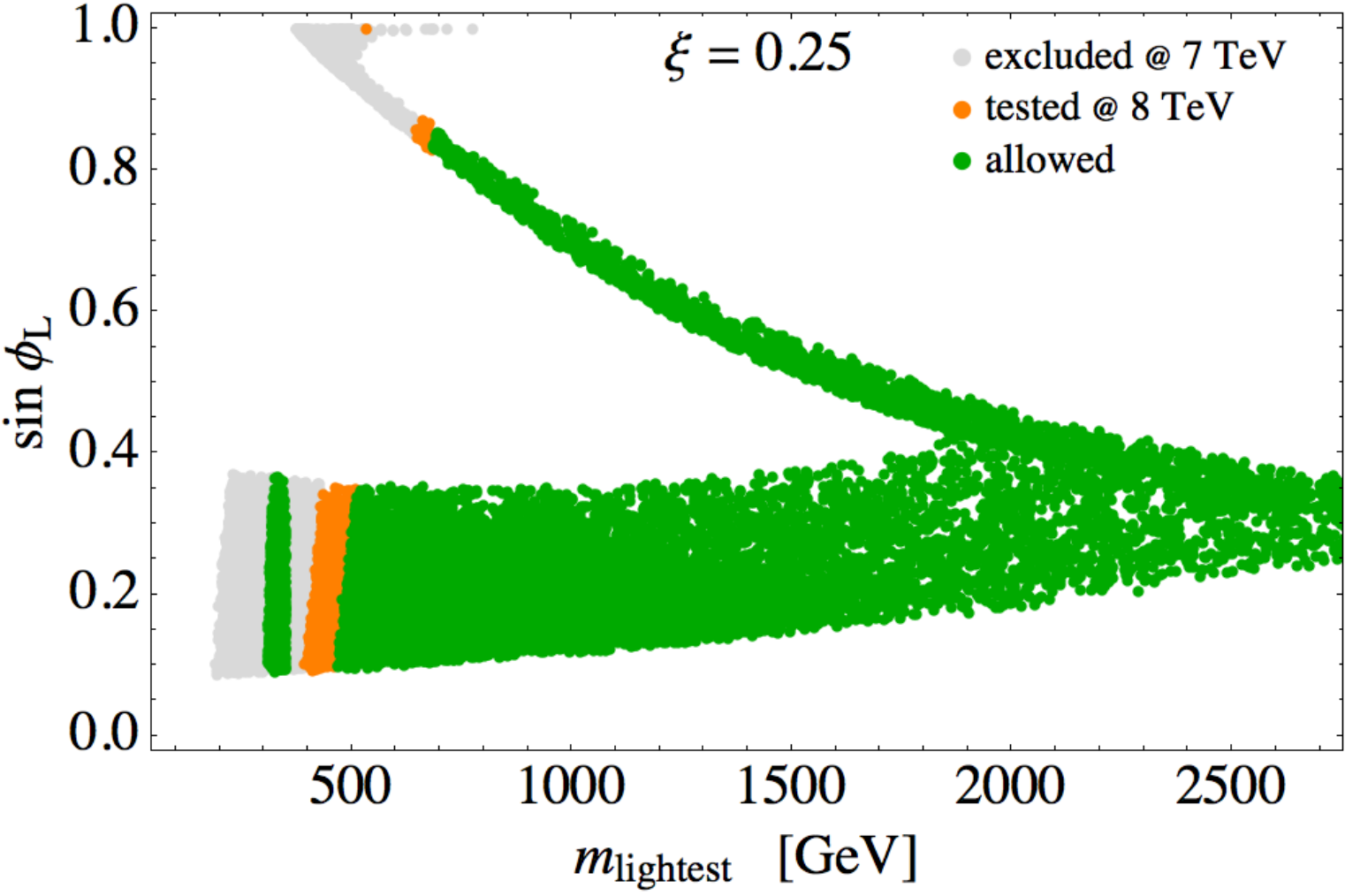} \hspace*{0.5cm}
\includegraphics[width=.47\textwidth]{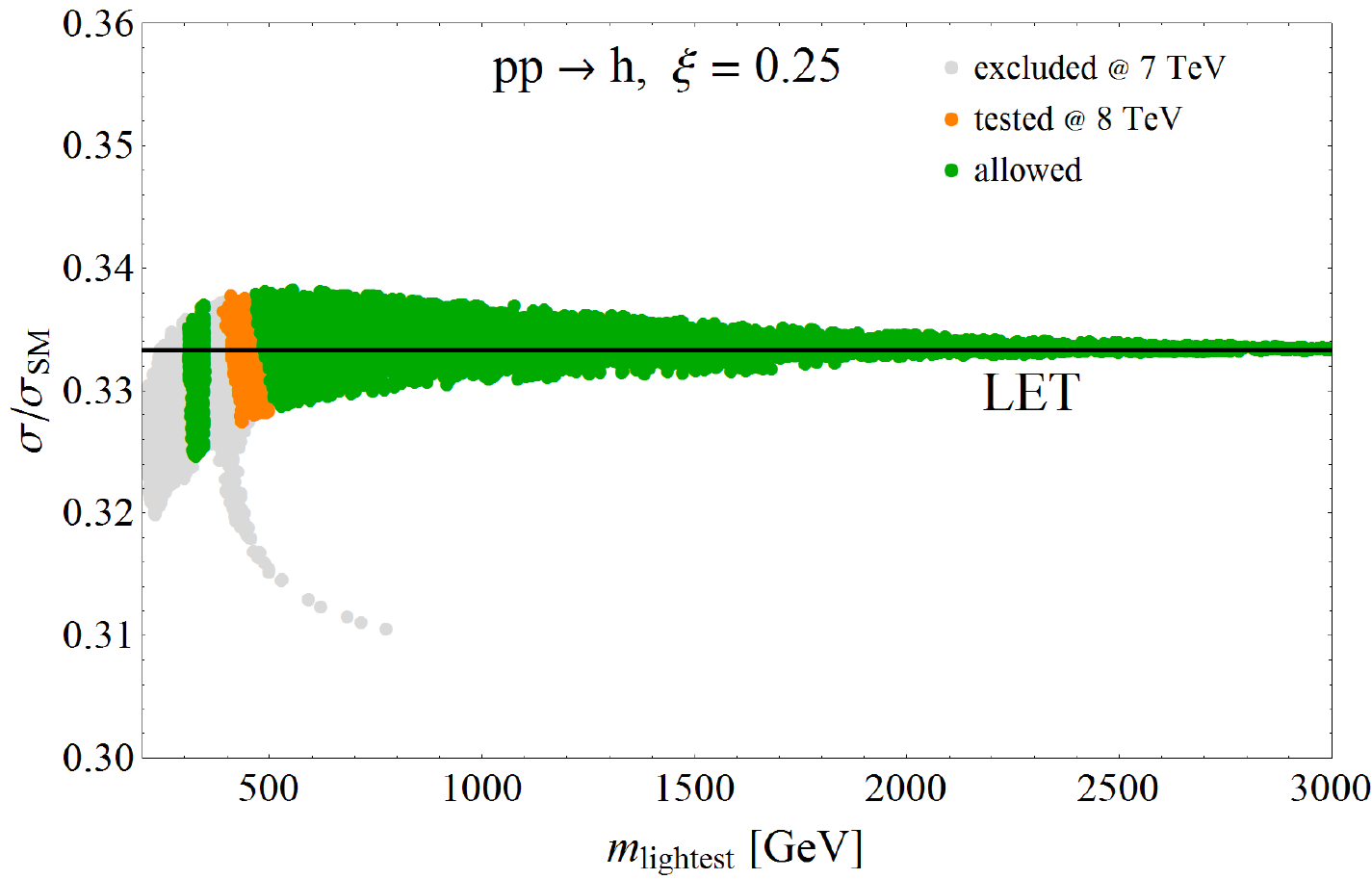}
	\caption{Left: Parameters passing the $\chi^2$-test of
          electroweak precision observables, displaying  the
          compositeness of the left-handed top versus the mass of the
          lightest top partner, for $\xi = 0.25$. 
Right: The MCHM$_5$ cross section for single Higgs production
  through gluon fusion (including the exact dependence on top and
  heavy fermion masses), normalized to the SM cross section (computed
  retaining the $m_{t}$ dependence), as a function of the mass of the
  lightest fermion resonance $m_{\rm{lightest}}$ for
  $m_{h}=125\,\mathrm{GeV}$ and $\xi=0.25$. The cross
  section ratio computed with the LET, Eq.~\eqref{eq:singlelet}, is shown
  as a black line.
          The points in light grey do not pass the
          direct collider constraints, points in orange/medium gray
          pass the 7~TeV constraints but are tested by the LHC
          running at 8~TeV with an integrated luminosity of \mbox{$15
            \,\mathrm{fb}^{-1}$}. Figures from Ref.~\cite{Gillioz:2012se}.}
	\label{fig:scanmchm5}
\end{figure*}

The strongest experimental constraints on the model come from the
electroweak precision measurements at the $Z$ pole mass at LEP. The
modified Higgs 
couplings to $W$ and $Z$ bosons induce logarithmically divergent
contributions to the $S$ and $T$ parameters, which are cut-off by the
mass $m_\rho$ of the first composite vector resonance
\cite{Contino:2010rs}. Another BSM effect is the direct contribution
of the vector $\rho$ and axial-vector $a$ resonances to the $S$
parameter. Finally the top partners give loop contributions both to
the $T$ parameter and the $Zb\bar{b}$ vertex
\cite{Agashe:2005dk,Lodone:2008yy,Gillioz:2008hs,Anastasiou:2009rv}. 
Figure~\ref{fig:scanmchm5} (left) from Ref.~\cite{Gillioz:2012se} shows the
points from a scan over the parameter range of the model after
applying a $\chi^2$ test, which assesses the agreement of the model
with the experimental data. These are the latest EW
precision data (EWPD) at that time, and the
constraint $|V_{tb}| > 0.77$ \cite{Group:2009qk} was applied, which also was the
then most up-to-date value. The results are displayed for the
left-handed compositeness angle $\phi_L$ versus the mass of the
lightest top partner, for the compositeness parameter
$\xi = 0.25$. The constraints on the parameter space from the EWPD can be
significantly relaxed by extending the fermion sector of the model
\cite{Anastasiou:2009rv,Gillioz:2013pba}.

Further constraints are due to flavor physics. In composite Higgs
models there arise generically four-fermion operators contributing to
flavor-changing (FC) processes and to electric dipole moments. The
constraints depend on the exact flavor structure of the model and can
be avoided in case of minimal flavor violation (MFV)
\cite{Redi:2011zi}. While both FC processes and electric dipole
moments are inhibited in this case, the MFV assumption requires a
large degree of compositeness also for light quarks, so that they have 
sizeable couplings to the strong sector resonances and lead to a
different phenomenology \cite{Delaunay:2013iia}. Dijet searches put
constraints on the up and down quarks
\cite{ATLAS:2012pu,Chatrchyan:2013muj}, while the second 
generation quarks are practically not constrained
\cite{DaRold:2012sz}. An alternative approach is to treat the top
quark differently than the light quarks \cite{Redi:2012uj}. The 
flavor bounds can still be satisfied, and since the first two
generations are mostly elementary the constraints from EWPT and
searches for compositeness are relaxed. In this case both the left-handed
and right-handed top can be composite. We do not assume a specific 
flavor model here. Additional discussions of flavor constraints on
composite Higgs models can be found {\it e.g.}~in
Ref.~\cite{Vignaroli:2012si}.

Another restriction of the model arises from direct searches for new
vector-like fermions by ATLAS and CMS. Based on the experimental results from the
searches for pair-produced fermions with subsequent decay into the
final states $WbWb$, $ZtZt$ and $WtWt$ given in
\cite{cmsfermsearch,atlasfermsearch,cdffermsearch}, exclusion limits
from direct searches have been included in
Figure~\ref{fig:scanmchm5} (left). As fermion pair production is a QCD
process it only depends on the mass $M_\psi$ of the heavy fermion
$\psi$. The constraint in a specific final state $X$ then reads,
\begin{equation}
\sigma_{\rm QCD}(pp\to \psi\overline{\psi})\times \mathrm{BR}(\psi\to
X)^{2}\leq \sigma_{\rm exp}\,, 
\end{equation}
where $\sigma_{\rm exp}$ is the upper bound on the cross section, as given
by the experiment for each value $M_\psi$. The QCD pair production
cross sections have been obtained at approximate NNLO
\cite{Aliev:2010zk}. The points that pass the direct search limits
from the 7~TeV run are shown in green in Fig.~\ref{fig:scanmchm5} (left). The
orange points are tested by LHC8. 

The gluon fusion cross section can be computed applying the LET and
can be cast into the form
\begin{equation}
\frac{\sigma(pp\to h)}{\sigma(pp\to h)_{\rm SM}} = 
\left(\frac{1-2\xi}{\sqrt{1-\xi}}\right)^{2}\,. \label{eq:singlelet}
\end{equation}
This result is valid to all orders in $\xi$. It depends purely on the
Higgs non-linearities and is independent of the details of the fermion
spectrum. When retaining the full mass dependence on the other hand,
corrections alter the cross section and are expected to be of the
order of at most a few percent. This is confirmed by the full
computation of the cross section taking into account the exact
dependence on the fermion masses, as can be inferred from
Fig.~\ref{fig:scanmchm5} (right).
%
%
%
\begin{figure}[t]
\begin{center}
\includegraphics[width=0.95\linewidth]{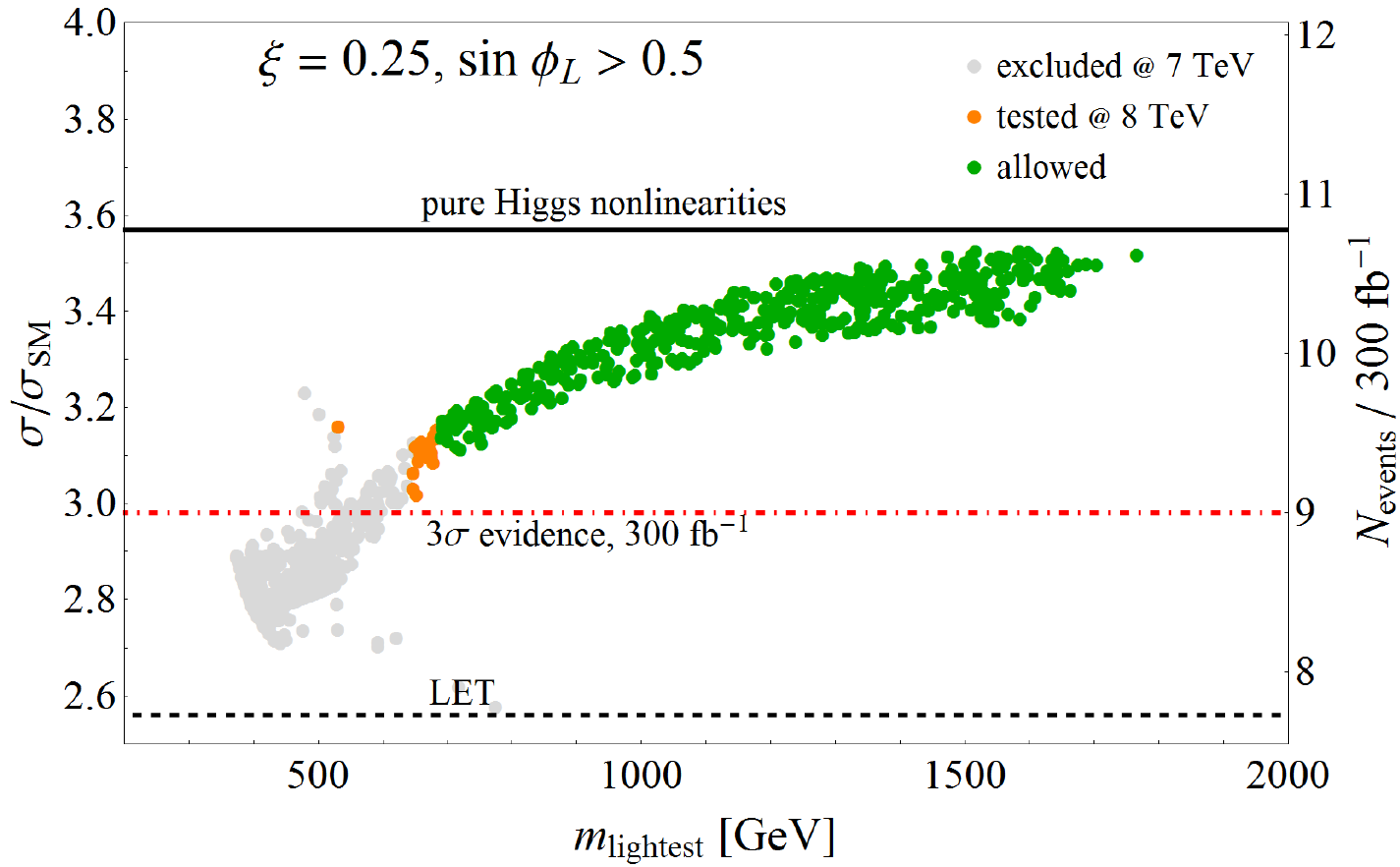} 
\vspace*{-0.6cm}
\end{center}
\caption{The double Higgs production cross section through gluon
  fusion normalized to the SM as function of the mass of the lightest
  heavy top partner, for $m_h=125$~GeV and $\xi=0.25$. Green/dark gray
  (gray) dots denote points 
  which pass (do not pass) the applied constraints, orange/fair gray
  points are tested by LHC8.  The lightest top
  partner is $X^{2/3}$. The black solid (dashed) line corresponds to 
  the result in the limit of heavy top partners keeping the full top
  mass dependence  (to the LET result). The expected number of events in the
  $hh\to b\bar{b}\gamma\gamma$ final state after all cuts at LHC14
  with $L=300\,\mathrm{fb}^{-1}$ is also shown, along with the
  $3\sigma$ evidence threshold (dot-dashed line), see
  Ref.~\cite{Gillioz:2012se}.} 
\label{fig:doublefull}
\end{figure}
The figure displays the cross section for single Higgs production
through gluon fusion including new fermionic resonances, normalized to
the SM cross section with the full top mass dependence, as a function
of the mass of the lightest resonance. The QCD $K$-factors, {\it
  i.e.}~the ratios between the QCD corrected cross section and the LO
result, cancel out under the assumption that the higher order
corrections are the same in both cases. In green are shown the points that pass
the EWPTs, gray and orange points do not satisfy the collider
bounds. The agreement between the full result and the prediction using
the low-energy theorem, shown by the black line, confirms that the
cross section for single Higgs production is almost independent of the
details of the spectrum and is fixed only by the value of $\xi$. For
heavy fermion partners the sensitivity to the composite couplings
practically vanishes. The LET hence provides a very accurate cross
section for single Higgs production for any spectrum of the heavy
fermions. The figure furthermore shows that in the MCHM5 single Higgs
production is suppressed compared to the SM for $\xi =0.25$. This can
be understood from the LET result given in Eq.~(\ref{eq:singlelet}).

Applying the LET to double Higgs production results in the partonic cross
section
\begin{equation}
\hat{\sigma}_{gg\to hh} = \frac{G_F^2 \alpha_s^2 (\mu) \hat{s}}{128 (2\pi)^3}
\, \frac{1}{9}\,\sqrt{1-\frac{4m_{h}^{2}}{\hat{s}}} \, C_{\textrm{LET}}^2 (\hat{s}) \;,
\label{eq:partoniccxn}
\end{equation} 
where $G_F$ denotes the Fermi constant and $\alpha_s$ the strong 
coupling constant evaluated at the scale $\mu$. The amplitude in the
limit of heavy loop particle masses at all orders in $\xi$ is given by
\begin{equation} 
\label{C in mchm5}
C^{\scriptsize \mbox{LET}}_{\rm MCHM5} (\hat{s}) = \frac{3m_h^2}{\hat{s}-m_h^2}
\left(\frac{1-2\xi}{\sqrt{1-\xi}}\right)^{2} -\frac{1}{1-\xi} \;.
\end{equation}
As in single Higgs production the LET cross section for Higgs pair
production is insensitive to the details of the heavy fermion spectrum
and fixed uniquely by $\xi$. 
In contrast to single Higgs
production, for Higgs pair production the cross section is enhanced
compared to the SM. In the SM Higgs pairs are produced through Higgs
bosons coupling to gluons via triangle loops and boxes. The related
amplitudes interfere destructively. In the MCHM5 the two contributions
are modified by $((1-\xi)/\sqrt{1-\xi})^2$, and there is an 
additional diagram with the two-Higgs two-fermion coupling
proportional to $\xi$, which can hence have order one effects and thus
govern the total cross section. 

Figure~\ref{fig:doublefull} shows the double Higgs production process
normalized to the SM cross section as a function of the lightest top
partner mass. The points result from a scan over
the parameter range of the model. In the computation of the cross
sections the full mass dependence of the loop particles has been taken
into account. In the black solid line the full mass dependence on the
top quark mass has been kept while the heavy fermion partners have
been integrated out. The dashed line displays the result in the LET
approximation, {\it i.e.}~sending all loop particle masses to
infinity. The color code is the same as in
Fig.~\ref{fig:scanmchm5}. The figure shows that there is a sizeable
dependence of the cross section on the spectrum of the heavy
fermions. We have $2.7 \lsim \sigma/\sigma_{\rm SM} \lsim 3.7$. The LET
cross section and the one in the limit of heavy partners on the other
hand only depend on $\xi$. However, the LET approximation considerably
underestimates the full result. Note that in the figure the cross
section is consistently normalized to the SM result for $m_t \to
\infty$. The result in the limit of heavy partners and keeping the
full top mass dependence on the other hand, overestimates the cross
section for masses of the lightest fermion partner below about
1~TeV. This is the region that is compatible with a 125~GeV Higgs
boson. For larger masses of the fermion partners, the cross section
approaches the result retaining only the top-loop contribution, as
expected. Higher-order QCD corrections have not been taken into
account in the figure. Due to the additional two-Higgs two-fermion
diagram and the box diagram with different loop particle masses they
cannot be taken over from the SM. For heavy loop particle masses the
corrections should not be too different from the SM case though, so
that they approximately cancel out in the ratio of the cross sections.

\begin{figure}
\begin{center}
   \includegraphics[scale=0.3, angle=-90]{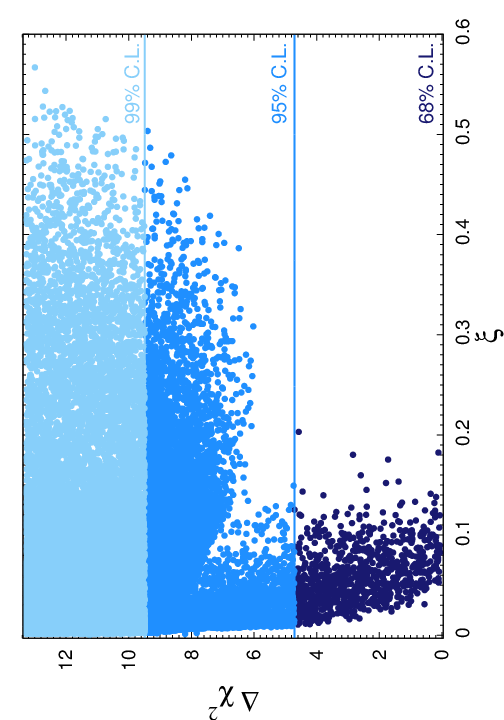}
\caption{Parameters passing the $\chi^2$ test of electroweak precision
  observables, fulfilling in addition $|V_{tb}| > 0.92$, for a scan
  over the parameter space. Details can be found in
  Ref.~\cite{Gillioz:2013pba}. Dark blue: $68\%$
  C.L. region, medium blue: $95\%$ C.L. region and light blue: $99\%$
  C.L. region, in the $\Delta\chi^2$ versus $\xi$ plane.}
\label{fig:bcompscan}
\end{center}
 \end{figure}

So far we have only treated the case of composite top quarks. With the bottom
quark being the next-heaviest quark a sizeable mixing with the strong
sector can be expected also for the bottom quark. Due to the small
bottom mass, the LET cannot be applied any more and the loop-induced
Higgs couplings to gluons are expected to depend on the resonance
structure of the strong sector, with significant implications for the
Higgs phenomenology
\cite{Gillioz:2013pba,Azatov:2011qy,Delaunay:2013iia}. The
effects of composite bottom quarks on composite Higgs models and
the LHC Higgs phenomenology can be studied by embedding the fermions
in the {\bf 10}. This is the smallest possible representation of
$SO(5)$ that allows to include partial compositeness for the bottom
quarks and at the same time is compatible with EWPTs by implementing
custodial symmetry. The bottom quark mass is in this case not introduced at
hoc any more but generated through the mixing with the strong
sector. The {\bf 10} leads to a larger spectrum of new heavy fermions
compared to the previous case with composite top quarks only. The new
vector-like fermions transform under the $SU(2)_L \times SU(2)_R$ and
are given by $u$, $u_1$, $t_4$ and $T_4$ with electric charge 2/3,
$d$, $d_1$ and $d_4$ with charge -1/3 and finally $\chi$, $\chi_1$ and
$\chi_4$ with charge 5/3. The model is strongly constrained by
EWPTs, in particular there are one-loop contributions to the
non-oblique corrections to the $Zb_L \bar{b}_L$ vertex due to the
partial compositeness of the bottom quark. This coupling has been
measured very precisely and agrees with the SM prediction at the
sub-percent level. It is safe from large corrections for the fermion
embedding in the {\bf 10} representation, where $b_L$ belongs to a
bi-doublet of $SU(2)_L \times SU(2)_R$ and the $SO(4)$ is enlarged to
$O(4)$. The general formulae for the contributions of new bottom
partners in the loop corrections to 
$Zb_L b_L$ have been derived in \cite{Gillioz:2013pba} and can be
applied to other models with similar particle content. In order to
test the viability of the model a scan over its parameter space has
been performed, setting the top and bottom quark masses to
$m_t=173.2$~GeV and $m_b=4.2$~GeV, respectively, and the Higgs boson
mass to $m_h=125$~GeV. A $\chi^2$ test is performed including the
constraints from EWPD, and only those points have been retained that
also fulfil $|V_{tb}| > 0.92$ \cite{Chatrchyan:2012ep}. The result is
shown in Fig.~\ref{fig:bcompscan} in the $\Delta \chi^2 \equiv \chi^2
- \chi^2_{\rm min}$ versus $\xi$ plane. The smallest $\Delta \chi^2$
values are obtained for $0.01 < \xi < 0.2$. High $\xi$ values lead to
large $\Delta \chi^2$, corresponding to a compatibility with EWPTs at
99\% confidence level (C.L.). The positive fermionic contributions to
the $T$ parameter play an important role here for the compatibility as
they drive back the $T$ value into the region in accordance with the EW
precision data. 
\begin{table*}
\centering
\begin{tabular}{c|c|c|c||c|c|c}\hline
&\multicolumn{3}{|c||}{$|V_{tb}|>0.92$}&\multicolumn{3}{|c}{ $|V_{tb}|$ in
$\chi^2$}\\\cline{2-7}
Experiment &$\xi$&$\chi^2/n$&$\chi^2_n$ &$\xi$&
$\chi^2/n$&$\chi^2_n$\\\hline
\multirow{2}{*}{ATLAS}&0.105&8.06/9&0.90&
 0.096&12.34/10&1.23\\\cline{2-7}
& 0.0 &17.54/13&1.35& 0.0 &17.73/14&1.25\\\hline
\multirow{2}{*}{CMS} &0.057&5.22/10&0.52&0.055&6.36/11&0.58\\\cline{2-7}
 &0.0& 9.90/14&0.71& 0.0& 10.09/15&0.67\\\hline
\end{tabular}
\caption{Global $\chi^2$ results for the best fit point taking into
  account EWPT and the Higgs results for ATLAS and CMS,
  respectively: {\it Left:} For parameter points which fulfil
  $|V_{tb}| >0.92$. {\it Right:} When including the measured
  value of $|V_{tb}|$ in the $\chi^2$ test.
  The lines for $\xi=0.0$ list for comparison the SM values. 
  The number of degrees of freedom $n$ are counted naively as the
  difference between the number of observables and the number of
  parameters in the model, and $\chi^2_n \equiv \chi^2/n$. Table from
  \cite{Gillioz:2013pba}.} 
\label{tab:bestfitbquark}
\end{table*}
The SM limit is obtained for $\xi \to 0$ and $M_{10} \to \infty$,
where $M_{10}$ sets the scale for the top and bottom partner
masses. Due to the restriction of the scan to $M_{10} \le 10$~TeV it
is not contained in the plot. 

The LHC Higgs search data put further constraints on the model. The
inclusion of heavy bottom and top quarks leads to changes in the Higgs
Yukawa couplings and new heavy fermions in the loop-induced production
and decay processes. In particular, in contrast to the heavy top
partner case, in the dominant gluon fusion production process due to
the mixing with bottom partners the LET cannot be applied any more, so
that the cross section now depends on the details of the model. Also
the compatibility with the direct searches for new vector-like fermions
performed by ATLAS \cite{atlassearch} and CMS
\cite{cmssearch1,Chatrchyan:2013uxa} has to be tested. Flavor physics
further constrains the model. As this depends on the exact flavor
structure of the model which is not specified here, these constraints
are not taken into account in this investigation. For a
random scan over the parameter range a $\chi^2$ test is performed
taking into account EWPTs, Higgs data and the experimental results on
$V_{tb}$. The results from direct fermion searches are taken into
account by discarding all points that lead to masses below the
exclusion limits. In Table~\ref{tab:bestfitbquark} the $\chi^2$
values of the best fit points are reported together with those for the
SM for comparison. The constraint from $V_{tb}$ has been included in
two different ways. Either all points with $|V_{tb}|> 0.92$ are
rejected or the best fit value given by the experiments is included in
the $\chi^2$ test. The global $\chi^2$ is increased when including
$V_{tb}$, in particular in the composite Higgs model. The table shows
that the CMS \cite{CMSdata} data are better described than the ATLAS
data \cite{ATLASdata}. The best fit points are obtained for $\xi
\approx 0.1$ for ATLAS and for $\xi \approx 0.05$ for CMS. In the
composite Higgs model their $\chi^2$ is slightly smaller than in the
SM, due to the larger number of free parameters. An estimate of the
relative goodness of the fit is given by $\chi^2_n \equiv \chi^2 /n$,
where $n$ denotes the number of degrees of freedom. 


Our discussion shows that composite Higgs models, although
challenged by constraints from EWPTs, flavor physics, direct searches
for new fermions and the light Higgs mass, are still viable extensions
beyond the SM Higgs. They provide an explanation of EWSB based on
strong dynamics and hence an alternative to weakly coupled models as
{\it e.g.}~supersymmetry, which shall be discussed in the following.

\section{Supersymmetric Higgs Models}
\label{sec:susy}
Supersymmetric theories 
\cite{Volkov:1973ix,Wess:1974tw,Fayet:1976et,Fayet:1976cr,Fayet:1977yc,Fayet:1979sa,Farrar:1978xj,Witten:1981nf,Nilles:1983ge,Haber:1984rc,Sohnius:1985qm,Gunion:1984yn,Lahanas:1986uc,Dimopoulos:1981zb,Sakai:1981gr}
provide a natural
solution to the hierarchy problem by introducing a new symmetry
between fermionic and bosonic degrees of freedom. They are among the most
extensively studied BSM extensions. In order to ensure supersymmetry
(SUSY) and an anomaly-free theory two complex Higgs doublets have to
be introduced, $H_u$ to provide masses to the up-type fermions, and
$H_d$ to ensure masses for the down-type fermions. In sections~\ref{sec:mssm} 
and \ref{sec:nmssm} we will discuss Higgs physics in two popular supersymmetric models, the minimal (MSSM)~\cite{mssm} and next-to-minimal supersymmetric model (NMSSM)~\cite{nmssm}, respectively. 

\subsection{Associated heavy quark Higgs production in the MSSM}
\label{sec:mssm}
After electroweak symmetry breaking, three of the eight degrees of freedom contained in 
$H_u$ and $H_d$ are absorbed by the $Z$ and $W$ gauge bosons, leaving five 
elementary Higgs particles in the MSSM. These consist of two CP-even neutral (scalar) particles
$h,H$, one CP-odd neutral (pseudoscalar) particle $A$, and two charged
bosons $H^\pm$.  At leading order the MSSM Higgs sector is fixed by two
independent input parameters which are usually chosen as the
pseudoscalar Higgs mass $M_A$ and $\tgb=v_u/v_d$, the ratio of the
vacuum expectation values of $H_u$ and $H_d$, respectively. Including the one-loop and dominant two-loop
corrections the upper bound of the light scalar Higgs mass is $M_h\lsim
135$~GeV \cite{mssmrad} for supersymmetric mass scales up to about a TeV.

An important property of the bottom Yukawa couplings is their enhancement for large values of $\tgb$. 
The top Yukawa couplings, on the other hand, are suppressed for large $\tgb$
\cite{schladming}, unless the light (heavy) scalar Higgs mass is close to its 
upper (lower) bound, where their couplings become SM-like. The
couplings of the various neutral MSSM Higgs bosons to fermions and gauge
bosons, normalized to the SM Higgs couplings, are listed in
Table~\ref{tb:hcoup}, where the angle $\alpha$ denotes the mixing angle
of the scalar Higgs bosons $h,H$.
\begin{table}[h]
\renewcommand{\arraystretch}{1.25}
\begin{center}
{\small \begin{tabular}{lc|ccc} 
\multicolumn{2}{c|}{$\phi$} & $g^\phi_u$ & $g^\phi_d$ &  $g^\phi_V$ \\
\hline
SM & $H$ & 1 & 1 & 1 \\ \hline
MSSM & $h$ & $\cos\alpha/\sin\beta$ & $-\sin\alpha/\cos\beta$ &
$\sin(\beta-\alpha)$ \\ & $H$ & $\sin\alpha/\sin\beta$ &
$\cos\alpha/\cos\beta$ & $\cos(\beta-\alpha)$ \\
& $A$ & $ 1/\tgb$ & $\tgb$ & 0 
\end{tabular}}
\renewcommand{\arraystretch}{1.2}
\caption[]{\label{tb:hcoup} \it MSSM Higgs couplings to 
$u$- and $d$-type fermions
and gauge bosons [$V=W,Z$] relative to the SM couplings.}
\end{center}
\end{table}

The negative direct searches for neutral MSSM Higgs bosons at LEP2 have lead to
lower bounds of $M_{h,H} > 92.8$ GeV and $M_A > 93.4$ GeV~\cite{lep2}. The LEP2 results 
also exclude the range $0.7 < \tgb < 2.0$ in the MSSM, assuming a SUSY
scale $M_{\rm SUSY}=1$~TeV \cite{lep2}. MSSM Higgs-boson searches have continued at 
the $p\bar{p}$ collider Tevatron, see e.g.\ \cite{Aaltonen:2012zh}, 
 and form the central part of the 
current and future physics program of the LHC \cite{atlas_cms_tdrs}.
The present LHC searches at 7 and 8 TeV c.m.~energy have excluded
parts of the MSSM 
parameter space for large values of $\tgb$ \cite{lhcmssm}. However, the recent
discovery of a resonance with a mass near 125~GeV
\cite{:2012gk,:2012gu} is a clear indication for the existence of a SM or beyond-the-SM Higgs boson. 
While the properties of the new particle, as determined so far, are consistent with those predicted 
within the SM, the bosonic state at 125~GeV can also be 
interpreted as a supersymmetric Higgs boson. 

Neutral MSSM Higgs boson production at  the LHC is dominated 
by gluon fusion, $gg\to h/H/A$, and by the associated production of a Higgs boson with bottom quarks. 
Gluon fusion is most significant at small and moderate $\tgb$. At large values of  $\tgb$, however, bottom--Higgs associated 
production 
\begin{equation}\label{eq:bbh}
q\bar q/gg\to b\bar b + h/H/A
\end{equation}
constitutes the dominant Higgs-boson production process. Here, we shall focus on precision 
calculations for bottom--Higgs associated production at the LHC. We refer the reader to 
Refs.~\cite{Dittmaier:2011ti, Dittmaier:2012vm,Heinemeyer:2013tqa} and references therein for a 
discussion of higher-order calculations for gluon fusion.

The NLO QCD
corrections to Higgs-bottom associated production, Eq.~(\ref{eq:bbh}), can be inferred from the analogous calculation involving top
quarks \cite{tthnlo,tthnlo2}. However, they turn out to be numerically enhanced \cite{bbhnlo} by large logarithmic contributions 
from the phase space region with final-state bottom quarks at small small transverse momenta. 
Those logarithms can be resummed by introducing bottom-quark densities in 
the proton~\cite{Barnett:1987jw}
and by applying the standard DGLAP evolution. In this so-called 
five-flavor scheme (5FS) the leading-order process is  
\cite{bbh5lo}
\begin{displaymath}
b\bar b\to h/H/A\,,
\end{displaymath}
where the transverse momenta of the incoming bottom quarks, their
masses, and their off-shellness are neglected at LO.  The
NLO~\cite{bbh5nlo} and NNLO \cite{bbh5nnlo} QCD corrections as well as
the SUSY--electroweak corrections \cite{bbh5elw} to these
bottom-initiated processes have been calculated. They are of moderate size, if the
running bottom Yukawa coupling is introduced at the scale of the
corresponding Higgs-boson mass.  The fully exclusive $gg\to b\bar b +
h/H/A$ process, calculated with four active parton flavors in a
four-flavor scheme (4FS), and the 5FS calculation converge at higher perturbative orders, and there is fair 
numerical agreement between the NLO 4FS and NNLO 5FS cross 
section predictions \cite{Dittmaier:2011ti,Dittmaier:2012vm,bbhnlo,bbhscale}.  
In Ref.~\cite{Harlander:2011aa} a scheme has been proposed to match the 4FS and 5FS cross sections 
(``Santander matching"). The scheme is based on the observation that the 4FS and 5FS calculations 
of the cross section are better motivated in the asymptotic limits of small and large Higgs masses, 
respectively, and combines the two approaches in such a way that they are given variable weight, depending on 
the value of the Higgs-boson mass.

If both bottom jets accompanying the Higgs boson
in the final state are tagged, one has to rely on the fully exclusive
calculation for $gg\to b\bar b + h/H/A$. For the case of a single
$b$-tag in the final state the corresponding calculation in the 5FS
starts from the process $bg\to b+h/H/A$ with the final-state bottom
quark carrying finite transverse momentum. 
The NLO QCD, electroweak, and NLO SUSY-QCD corrections to this process have been calculated~\cite{bh5elw}.

State-of-the-art predictions as well as estimates of the corresponding
parametric and theoretical uncertainties have been provided by the 
LHC Higgs Cross Section Working Group both for 
total~\cite{Dittmaier:2011ti} and differential~\cite{Dittmaier:2012vm,Heinemeyer:2013tqa}
cross sections.

Recently,  the 4FS calculations for heavy quark plus Higgs associated production at the LHC 
have been improved by including the full SUSY-QCD corrections within the MSSM
and separating the dominating universal part in terms of effective Yukawa
couplings~\cite{Dittmaier:2014sva}. The SUSY loop corrections modify the tree-level relation between the
bottom mass and its Yukawa coupling, which is enhanced at large
$\tgb$~\cite{deltab}. These corrections can be summed to all orders
by replacing the bottom Yukawa coupling coefficients of
Table~\ref{tb:hcoup} by \cite{Carena:1999py,Guasch:2003cv}
\begin{eqnarray}
\tilde g_b^h & = & \frac{g_b^h}{1+\Delta_b} \left(
1-\frac{\Delta_b}{\tga~\tgb}\right), \nonumber \\
\tilde g_b^H & = & \frac{g_b^H}{1+\Delta_b} \left(
1+\Delta_b \frac{\tga}{\tgb}\right), \nonumber \\
\tilde g_b^A & = & \frac{g_b^A}{1+\Delta_b} \left(
1-\frac{\Delta_b}{\stgb}\right),
\label{eq:dmb}
\end{eqnarray}
where
\begin{equation}
\begin{aligned}
\hspace*{-4mm}\Delta_b & =  \frac{C_F}{2}~\frac{\alpha_{\rm s}}{\pi}~m_{\sgl}~\mu~\tgb~
I(m^2_{\ssb_1},m^2_{\ssb_2},m^2_{\sgl}), \\
\hspace*{-3.5mm}I(a,b,c) & =  \frac{\displaystyle ab\log(a/b) +
bc\log(b/c)
+ ca\log(c/a)}{(a-b)(b-c)(a-c)} 
\end{aligned}
\end{equation}
with $C_F = 4/3$. Here, $\tilde{b}_{1,2}$ are the sbottom mass
eigenstates, and $m_{\tilde{g}}$ denotes the gluino mass.

To discuss the various contributions to the NLO SUSY-QCD corrections, we focus 
on $b\bar{b}H$ production as an example, and write the NLO cross section as 
\beq
\sigma^H_{\rm NLO} = \sigma^H_0\times (1+\delta^H_{\rm SUSY})\times
(1+\delta^H_{\rm QCD}+\delta^H_{\rm SUSY-rem}),
\label{eq:corrbbh}
\eeq
where $\sigma^H_0$
denotes the LO cross section evaluated with LO $\alpha_{\rm s}$ and PDFs,
with the Yukawa coupling parametrized in terms
of the running b-quark mass $\overline{m}_b(\mu)$, but
without resummation of the $\tgb$-enhanced terms.
The correction $\delta^H_{\rm SUSY}$ comprises the $\tgb$-enhanced
terms according to Eq.~(\ref{eq:dmb}), including their resummation.
The remainder of the
genuine SUSY-QCD corrections is denoted by $\delta^H_{\rm SUSY-rem}$.

\begin{figure}[ht]
\includegraphics[width=0.475\textwidth]{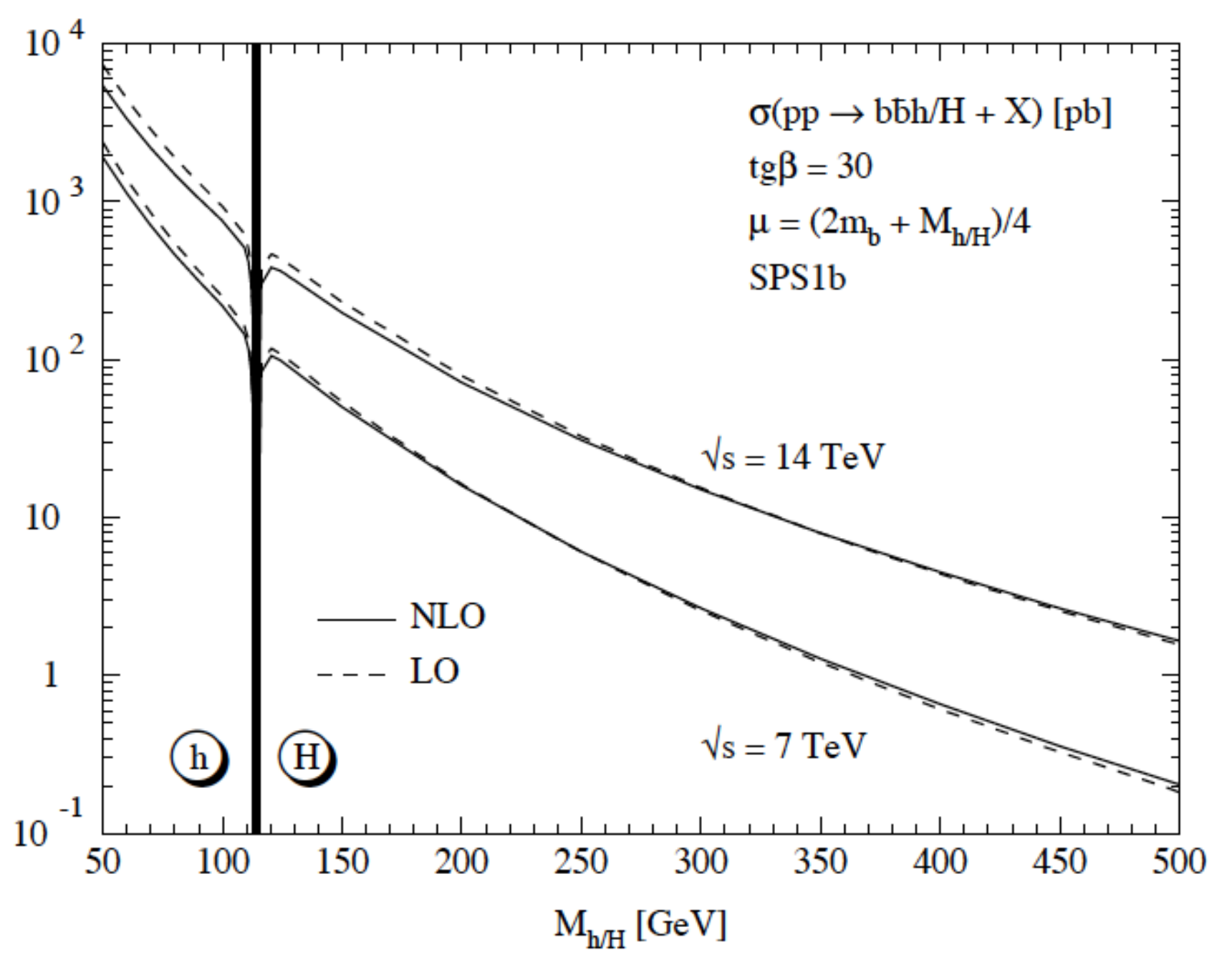}
\caption[]{\label{fg:cxn_bbh} \it SUSY-QCD corrected production cross
sections of light and heavy scalar MSSM Higgs bosons in association with
$b\bar b$ pairs for the Snowmass point SPS1b \cite{snowmass}. See Ref.~\cite{Dittmaier:2014sva} for details.}
\end{figure}

The total cross section for scalar Higgs-boson radiation off bottom quarks at the LHC ($\sqrt{S} = 14$\,TeV)
is shown in Fig.~\ref{fg:cxn_bbh}
for the MSSM benchmark scenario SPS1b \cite{snowmass} 
with $\tgb = 30$. The four-flavor MSTW2008 pdfs \cite{mstw4} have been used.  
In Table~\ref{tb:susy-qcd} we show the individual contributions to the
NLO cross section. 
\begin{table}
\renewcommand{\arraystretch}{1.2}
\begin{center}
{\small \begin{tabular}{l|cc|ccc} 
& $M_A$ & $M_H$ [GeV] & $\delta^H_{\rm QCD}$ & $\delta^H_{\rm SUSY}$ &
$\delta^H_{\rm SUSYrem}$ \\[1mm]
\hline
       & 100 & 113.9 & 0.27 & $-0.38$ & $0.3\times 10^{-4}$ \\
       & 200 & 200   & 0.39 & $-0.30$ & $5.8\times 10^{-4}$  \\
7 TeV\!\!  & 300 & 300    & 0.47 & $-0.30$ & $9.3\times 10^{-4}$  \\
       & 400 & 400    & 0.53 & $-0.30$ & $1.5\times 10^{-3}$  \\
       & 500 & 500   & 0.59 & $-0.30$ & $2.2\times 10^{-3}$  \\
\hline
       & 100 & 113.9 &  0.17 & $-0.38$ & $0.5\times 10^{-4}$ \\
       & 200 & 200   &  0.29 & $-0.30$ & $5.7\times 10^{-4}$  \\
14 TeV\!\! & 300 & 300    & 0.39 & $-0.30$ & $9.3\times 10^{-4}$  \\
       & 400 & 400   & 0.45 & $-0.30$ & $1.5\times 10^{-3}$  \\
       & 500 & 500   & 0.49 & $-0.30$ & $2.3\times 10^{-3}$  
\end{tabular}}
\caption[]{\label{tb:susy-qcd} \it Individual NLO corrections relative
to the LO cross section for $pp\to b\bar b H + X$, 
as defined in Eq.~(\ref{eq:dmb}),
are shown for the
LHC at two c.m.~energies (7~TeV and 14~TeV) for the Snowmass point SPS1b
\cite{snowmass}.}
\end{center}
\vspace*{-0.5cm}
\end{table}
The  pure QCD corrections $\delta^H_{\rm QCD}$, 
the $\tgb$-enhanced SUSY-QCD corrections $\delta^H_{\rm SUSY}$,
and the remainders of the SUSY-QCD corrections
$\delta^H_{\rm SUSY-rem}$ are defined according to Eq.~(\ref{eq:corrbbh}).
The moderate NLO corrections in this MSSM scenario result from a compensation 
of the large, positive QCD corrections and large, negative SUSY-QCD corrections.
The smallness of the SUSY-QCD remainder shows that the
full NLO SUSY-QCD corrections are approximated extremely well by the 
$\tgb$-enhanced terms. 

Let us briefly comment on heavy charged Higgs production at the LHC, 
\begin{equation} 
pp\, \rightarrow\, tH^{\pm}(b)+X.
\end{equation}
In a two-Higgs doublet model of type II,  like the minimal supersymmetric extension of the SM,  
the Yukawa coupling of the charged Higgs boson $H^-$ to a top quark and bottom antiquark is given by 
\begin{equation}
g_{t\bar{b}H^-} = \sqrt{2}\left(\frac{m_t}{v}P_R\cot\beta + \frac{m_b}{v}P_L\tan\beta\right),
\label{eq:tbhcoupling}
\end{equation}
where $v = \sqrt{v_u^2+v_d^2} = (\sqrt{2}G_F)^{-\frac12}$ is the Higgs vacuum expectation value in the 
SM, with the Fermi constant $G_F = 1.16637 \times 10^{-5}\;\mbox{GeV}^{-2}$, and $P_{R/L} = (1\pm\gamma_5)/2$ are the chirality projectors. 

As in the case of neutral Higgs production with bottom quarks, the cross section for $pp\, \rightarrow\, tH^{\pm}(b)+X$ 
can be calculated in the four- or five-flavor schemes. Next-to-leading order predictions for heavy charged Higgs boson
production at the LHC in a type-II two-Higgs-doublet model 
have been made in the past in both 5F~\cite{Zhu:2001nt,Gao:2002is,Plehn:2002vy,Berger:2003sm,Kidonakis:2005hc,Weydert:2009vr} 
and 4F schemes~\cite{Peng:2006wv,Dittmaier:2009np}, including also electroweak corrections~\cite{Beccaria:2009my,Nhung:2012er}. 
In Ref.~\cite{Flechl:2014wfa}, the NLO-QCD predictions have recently been updated and improved by adopting a dynamical scale setting 
procedure for the 5FS~\cite{Maltoni:2012pa}.  A thorough account of all sources of theoretical uncertainties has been given, 
and a Santander-matched prediction~\cite{Harlander:2011aa} for the four- and five-flavor scheme calculations has been provided. 
Reference~\cite{Flechl:2014wfa} also includes results 
for a wide range of $\tan\beta$, which allows the comparison 
between theory and experiment for a large class of beyond-the-SM scenarios. 

\begin{figure}
\begin{center}
  \includegraphics[width=0.48\textwidth]{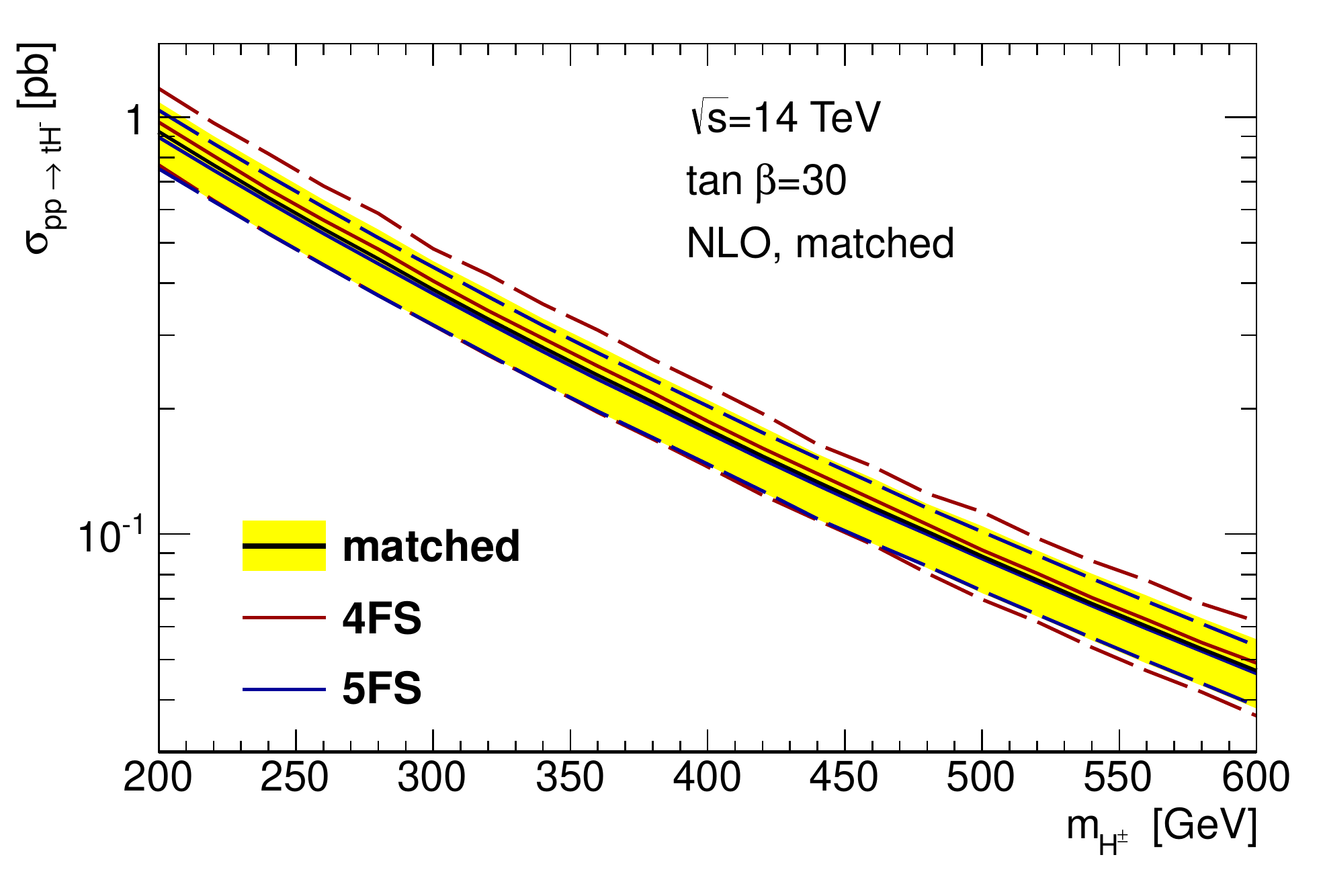}
\end{center}
        \caption{Santander-matched cross section and uncertainties for 
$pp \rightarrow tH^\pm + X$ at the LHC for 14\,TeV. 
The 4F and 5F scheme results as well as the combined values are shown, 
together with their total uncertainties.
}
	\label{fig:matched}
\vspace*{-0.2cm}
\end{figure}

The cross section and uncertainty for the results of the 4F and 5F scheme calculations and their combination 
for $\sqrt{s}=14$\,TeV are presented in Fig.~\ref{fig:matched}. 
The dynamical choice of the factorisation scale in the five-flavor scheme 
calculation significantly improves the agreement between the four- and five-flavor schemes. The overall uncertainty 
of the matched cross-section prediction is approximately 20--30\%, and includes the dependence on the renormalisation scale, the factorisation scale, the 
scale of the running bottom quark mass in the Yukawa coupling, as well as the input parameter uncertainties in the parton distribution functions, in the QCD coupling $\alpha_s$, and in the bottom quark mass. 

\subsection{Higgs Bosons in the NMSSM}
\label{sec:nmssm}
The NMSSM \cite{nmssm} extends the Higgs sector by
an additional singlet superfield $\hat{S}$. This entails seven Higgs bosons
after EWSB, which in the limit of the real NMSSM can be divided into
three neutral purely CP-even, two neutral purely CP-odd and two
charged Higgs bosons, and in total leads to five neutralinos. The
NMSSM allows for the dynamical solution of the $\mu$ problem \cite{Kim:1983dt}
through the scalar component of the singlet field acquiring a
non-vanishing vacuum expectation value. Furthermore, the tree-level mass value of
the lightest Higgs boson is increased by new contributions to the
quartic coupling so that the radiative corrections necessary to shift
the Higgs mass to $\sim 125$~GeV are less important than in the
MSSM allowing for lighter stop masses and/or mixing and in turn less
finetuning. The singlet admixture in the Higgs mass eigenstates
entails reduced couplings to the SM particles. Together with the
enlarged Higgs sector this leads to a plethora of interesting
phenomenological scenarios and signatures. It is obvious that on the
theoretical side the reliable interpretation of such BSM signatures
and the disentanglement of different SUSY scenarios as well as their
distinction from the SM situation require precise predictions of the
SUSY parameters such as masses and Higgs couplings to other Higgs
bosons including higher-order corrections. For the CP-conserving NMSSM
the mass corrections are available at one-loop accuracy
\cite{Ellwanger:1993hn,Elliott:1993ex,Elliott:1993uc,Elliott:1993bs,Pandita:1993hx,Ellwanger:2005fh,Degrassi:2009yq,Staub:2010ty,Ender:2011qh}, and
two loop results of O($\alpha_t \alpha_s + \alpha_b \alpha_s$) in the
approximation of zero external momentum have been given in
Ref.~\cite{Degrassi:2009yq}. Recently, first corrections beyond order
O($\alpha_t \alpha_s + \alpha_b \alpha_s$) have been given
in \cite{Goodsell:2014pla}. In the complex 
case the one-loop corrections to the Higgs masses have been given in
\cite{Ham:2001kf,Ham:2007mt,Ham:2001wt,Ham:2003jf,Funakubo:2004ka,Graf:2012hh} with the logarithmically enhanced
two-loop effects presented in \cite{Cheung:2010ba}. 
Quite recently we have provided the two-loop corrections to the Higgs
boson masses of the CP-violating NMSSM in the Feynman diagrammatic
approach with vanishing external momentum at O($\alpha_t \alpha_s$)
\cite{Muhlleitner:2014vsa}. 
Higher-order corrections to the trilinear Higgs self-coupling of the neutral NMSSM
Higgs bosons have been provided in \cite{Nhung:2013lpa}.

In the following the implications of the loop corrections in the real
and complex NMSSM to the Higgs boson masses, and for the CP-conserving
NMSSM, to the trilinear Higgs self-couplings shall be presented. To
set up the notation the NMSSM Higgs sector is briefly discussed. 
The Higgs mass matrix is derived from the NMSSM Higgs potential,
which is obtained from the superpotential, the soft SUSY breaking
terms and from the $D$-term contributions. Denoting the Higgs 
doublet superfields, which couple to the up- and down-type quarks, by
$\hat{H}_u$ and $\hat{H}_d$, respectively, and the singlet superfield by
$\hat{S}$, the scale invariant NMSSM superpotential is given by
\begin{eqnarray}
W_{\rm NMSSM} = W_{\rm MSSM} - \epsilon_{ij} \lambda \hat{S} \hat{H}^i_d
\hat{H}^j_u + \frac{1}{3} \kappa \hat{S}^3 \;,
\end{eqnarray}
with the $SU(2)_L$ indices $i,j=1,2$ and the totally antisymmetric
tensor $\epsilon_{12}=1$. While the dimensionless parameters $\lambda$
and $\kappa$ can be complex in general, in case of CP-conservation
they are taken to be real. In terms of the quark and lepton
superfields and their charge conjugates, indicated by the superscript
$c$, $\hat{Q}, \hat{U}^c, \hat{D}^c, \hat{L}, \hat{E}^c$, the MSSM
superpotential $W_{\rm MSSM}$ reads
\begin{eqnarray}
W_{\rm MSSM} &=& \epsilon_{ij} [y_e \hat{H}^i_d \hat{L}^j \hat{E}^c + y_d
\hat{H}_d^i \hat{Q}^j \hat{D}^c \nonumber \\
&& - y_u \hat{H}_u^i \hat{Q}^j \hat{U}^c] 
\label{eq:mssmsuperpot} \;,
\end{eqnarray}
where the flavor and generation indices have been suppressed. 
In general the Yukawa couplings $y_d,$ $y_u$ and $y_e$ in
the MSSM superpotential Eq.~(\ref{eq:mssmsuperpot}) are complex. 
Neglecting generation mixing, their phases can be reabsorbed by 
redefining the quark fields. The MSSM $\mu$-term as well as the
tadpole and bilinear terms of $\hat{S}$ are assumed 
to be zero. With the Higgs doublet and singlet component fields 
$H_u$, $H_d$ and $S$, the soft SUSY breaking NMSSM Lagrangian is given by
\begin{eqnarray}
\mathcal L_{\rm soft} &=& {\cal L}_{\rm soft}^{\rm MSSM} - m_S^2 |S|^2 +
(\epsilon_{ij} \lambda 
A_\lambda S H_d^i H_u^j \nonumber \\
&& - \frac{1}{3} \kappa
A_\kappa S^3 + h.c.) \;, \label{eq:nmssmsoft}
\end{eqnarray}
where ${\cal L}_{\rm soft}^{\rm MSSM} $ denotes the soft SUSY breaking MSSM Lagrangian,
\begin{eqnarray}
{\cal L}_{\rm soft}^{\rm MSSM} 
\hspace*{0.5cm} &\hspace*{-0.7cm}=& \hspace*{-0.3cm}-m_{H_d}^2 H_d^\dagger
H_d - m_{H_u}^2 
H_u^\dagger H_u -
m_Q^2 \tilde{Q}^\dagger \tilde{Q} \nonumber \\
&\hspace*{-2.3cm}-& \hspace*{-1cm} m_L^2 \tilde{L}^\dagger \tilde{L}
- \, m_U^2 \tilde{u}_R^*  
\tilde{u}_R - m_D^2 \tilde{d}_R^* \tilde{d}_R 
\nonumber \\
&\hspace*{-2.3cm}-& \hspace*{-1cm} m_E^2 \tilde{e}_R^* \tilde{e}_R 
- \, (\epsilon_{ij} [y_e A_e H_d^i
\tilde{L}^j \tilde{e}_R^* \nonumber \\
&\hspace*{-2.3cm}+& \hspace*{-1cm} y_d A_d H_d^i \tilde{Q}^j \tilde{d}_R^*
- \, y_u A_u H_u^i \tilde{Q}^j
\tilde{u}_R^*] + h.c.) \nonumber \\
&\hspace*{-2.3cm}-& \hspace*{-1cm} \frac{1}{2}(M_1 \tilde{B}\tilde{B} + M_2 \,
\tilde{W}_k\tilde{W}_k + M_3 \tilde{G}\tilde{G} + h.c.) \;.
\label{eq:mssmsoft}
\end{eqnarray}
Here $\tilde{B}$, $\tilde{W}_k$ ($k=1,2,3$) and $\tilde{G}$ are the
gaugino fields, and $\tilde{Q}=(\tilde{u}_L,\tilde{d}_L)^T$,
$\tilde{L}=(\tilde{\nu}_L,\tilde{e}_L)^T$, where the tilde indicates the 
scalar components of the corresponding quark and lepton
superfields. In the soft SUSY breaking NMSSM Lagrangian
Eq.~(\ref{eq:nmssmsoft}) the soft SUSY breaking mass parameters $m_X^2$
of the scalar fields $X=S,H_d,H_u,Q,U,D,L,E$ are real. The soft
SUSY breaking trilinear couplings $A_x$ ($x=\lambda,\kappa,d,u,e$) and
the gaugino mass parameters $M_1,M_2$ and $M_3$, however, are complex
in general but real in the CP-conserving case. Here, furthermore squark and
slepton mixing between the generations is neglected and soft SUSY breaking
terms linear and quadratic in the singlet field $S$ are set to
zero. In the expansion of the Higgs boson fields about the VEVs two
further phases, $\varphi_u$ and $\varphi_s$, appear,
\begin{eqnarray}
H_d &=& \begin{pmatrix}\frac{1}{\sqrt{2}}(v_d + h_d + i a_d) \\ 
                      h_d^- \end{pmatrix}~, \nonumber \\
H_u &=& e^{i \varphi_u} \begin{pmatrix} h_u^+ \\
            \frac{1}{\sqrt{2}}(v_u + h_u + i a_u) \end{pmatrix}~,
          \nonumber \\
S &=& \frac{e^{i \varphi_s}}{\sqrt{2}} (v_s + h_s + i a_s)~,
\label{Higgsdecomp}
\end{eqnarray}
describing the phase differences between the three VEVs $\langle
H_d^0 \rangle$, $\langle H_u^0 \rangle$ and $\langle S \rangle$. For
phase values $\varphi_u=\varphi_s= n\pi$, $n \in \mathbb{N}$, the
fields $h_i$ and $a_i$ ($i=d,u,s$) are the pure CP-even and CP-odd
parts of the neutral entries of $H_u$, $H_d$ and $S$. Exploiting the
freedom in the phase choice of the Yukawa couplings to set
$\varphi_{y_u}= - \varphi_u$ and assuming the down-type and charged
lepton-type Yukawa couplings to be real, the quark and lepton mass
terms yield real masses without any further phase transformation of
the corresponding fields. Expansion about the VEVs leads to the Higgs boson
mass matrix $M_{\phi\phi}$ that can be read off from the bilinear
neutral Higgs field terms in the Higgs potential. CP-violation
introduces a mixing between CP-even and CP-odd component 
fields, leading to a $6\times 6$ matrix in the basis $\phi
= (h_d, h_u, h_s, a_d, a_u, a_s)^T$, which can be expressed in terms of three $3
\times 3$ matrices $M_{hh}, M_{aa}$ and $M_{ha}$,
\begin{align}
M_{\phi\phi} = \begin{pmatrix} M_{hh}&  M_{ha}\\
                          M_{ha}^T &
                          M_{aa}\end{pmatrix} \;,
\label{eq:higgsmassmatrix}
\end{align}
where $M_{hh}$ and $M_{aa}$ are symmetric matrices, describing the
mixing among the CP-even components of 
the Higgs doublet and singlet fields and among the CP-odd components,
respectively. In the CP-conserving case the 
matrix $M_{ha}$, which mixes CP-even and CP-odd components,
vanishes. Due to the minimization conditions of the Higgs potential
$V$ ($\phi = h_d, h_u, h_s, a_d, a_u, a_s$),
\begin{eqnarray}
t_\phi \equiv \left. \frac{\partial V}{\partial \phi}
\right|_{\mbox{\scriptsize Min}} = 0 
\;,
\end{eqnarray}
in the tree-level Higgs sector only one linearly independent phase combination
$\varphi_y$ appears after applying the tadpole conditions for $\phi=a_d$
and $a_s$, 
\begin{eqnarray}
\varphi_y = \varphi_\kappa - \varphi_\lambda + 2\varphi_s - \varphi_u \;.
\end{eqnarray}
The tadpole condition for $a_u$ does not lead to a new linearly
independent condition. Hence the CP mixing due to $M_{ha}$ is governed
by $\sin\varphi_y$ at tree-level. In the real case, the CP-odd tadpole
conditions vanish and are thus automatically fulfilled.

In order to obtain the mass eigenstates first the $6 \times 6$
rotation matrix ${\cal R}^G$ is applied to separate the
would-be Goldstone boson field and then the matrix ${\cal R}$ to
rotate to the mass eigenstates $H_i$ ($i=1,...,5$), yielding a
diagonal mass matrix squared,
\begin{eqnarray}
(H_1,H_2,H_3,H_4,H_5,G)^T &\hspace*{-0.2cm}=\hspace*{-0.2cm}& \nonumber \\
&& \hspace*{-2cm} {\cal R}^D 
(h_d,h_u,h_s,a_d,a_u,a_s)^T \label{eq:cpR} \\
\mbox{diag} ((M_{H_1}^{(0)})^2,...,(M_{H_5}^{(0)})^2,0)
&\hspace*{-0.2cm}=\hspace*{-0.2cm}& {\cal R}^D  M_{\phi\phi} ({\cal
  R}^D)^T \label{eq:cpRm} 
\end{eqnarray}
with ${\cal R}^D \equiv {\cal R} {\cal R}^G$ and the superscript $(0)$
indicating tree-level masses. In the CP-conserving case the $6\times
6$ mass matrix $M_{\phi\phi}$ decomposes into $3 \times 3$ Higgs mass
matrices squared for the CP-even and CP-odd
component Higgs fields, $M_S^2$ and $M_A^2$, respectively. 
The squared mass
matrix $M_S^2$ is diagonalized through a rotation ${\cal
  R}^S$, yielding the CP-even mass eigenstates $H_i$ ($i=1,2,3$),
\begin{eqnarray}
 \begin{pmatrix} H_1,H_2,H_3 \end{pmatrix}^T =   {\cal R}^{S} \begin{pmatrix}
 h_d,h_u,h_s \end{pmatrix}^T, \label{eq:Srotation} \\
\text{diag}( (M_{H_1}^{(0)})^2,(M_{H_2}^{(0)})^2,(M_{H_3}^{(0)})^2) =
{\cal R}^S {M_S^2} 
({\cal R}^S)^T \; . 
\end{eqnarray}
The mass eigenstates are ordered by ascending mass, $M_{H_1}^{(0)} \le
M_{H_2}^{(0)} \le M_{H_3}^{(0)}$, where the superscript $(0)$ indicates the tree-level
mass values.  
In the CP-odd Higgs sector a first rotation ${\cal R}^G$ is applied to
separate the massless Goldstone boson $G$, followed by a 
rotation ${\cal R}^P$ to obtain the mass eigenstates $A_i \equiv A_1,
A_2, G$ ($i=1,2,3$), {\it cf.}~\cite{Ender:2011qh},
\begin{eqnarray}
 \begin{pmatrix} A_1,A_2,G \end{pmatrix}^T =   {\cal R}^{P}  {\cal
   R}^G \begin{pmatrix}  a_d,a_u,a_s \end{pmatrix}^T,  \\
\text{diag}((M_{A_1}^{(0)})^2,(M_{A_2}^{(0)})^2,0) = {\cal R}^P {\cal
  R}^G {M_A^2} 
({\cal R}^P {\cal R}^G)^T \label{eq:Protation} \; .
\end{eqnarray}

The parameters of the tree-level Higgs potential in the
CP-violating NMSSM are
\begin{eqnarray}
&& \hspace*{-1cm} 
m_{H_d}^2, m_{H_u}^2, m_S^2, g_1, g_2, v_u, v_d,  v_s, \nonumber \\
&& \hspace*{-1cm} \varphi_s, \varphi_u, 
 \Re\lambda, \Im\lambda,\Re A_{\lambda},\Im A_{\lambda}, \Re\kappa,
\Im\kappa, \Re A_{\kappa},\Im A_{\kappa}~.\label{eq:orgparset}
\end{eqnarray}
Some of the parameters are traded for more physical ones leading to 
the following parameter set
\begin{eqnarray}
&& \hspace*{-1cm} 
\underbrace{t_{h_d}, t_{h_u}, t_{h_s}, t_{a_d}, t_{a_s}, M_{H^\pm}^2,
  M_W^2, M_Z^2, e}_{\mbox{on-shell}},  \nonumber \\
&& \hspace*{-1cm} 
\underbrace{ \tan \beta,   v_s, \varphi_s, \varphi_u, \Re\lambda,
  \Im\lambda, \Re\kappa,  
\Im\kappa, \Re A_{\kappa}}_{\overline{\mbox{DR}}} \;.
\label{eq:defparset}
\end{eqnarray}
In the renormalization performed for the computation of the one-loop
corrected Higgs boson masses \cite{Graf:2012hh}, the first part of
parameters is defined via on-shell conditions. The tadpole parameters
are chosen such that the linear terms or the Higgs fields in the Higgs
potential also vanish at one-loop level. In a slight abuse of the
language the tadpole renormalization conditions are therefore also
called on-shell. The remaining parameters are interpreted as
$\overline{\mbox{DR}}$ parameters. An alternative renormalization
scheme uses the real part of $A_\lambda$ as a $\overline{\mbox{DR}}$
input parameter instead of the mass of the charged Higgs boson. In the
computation of the one-loop masses in CP-conserving NMSSM two further
renormalization schemes are applied, based on a pure on-shell,
respectively, a pure $\overline{\mbox{DR}}$ scheme
\cite{Ender:2011qh}. 
\begin{figure}[t!]
\begin{center}
\parbox{8.5cm}{\includegraphics[width=0.9\linewidth]{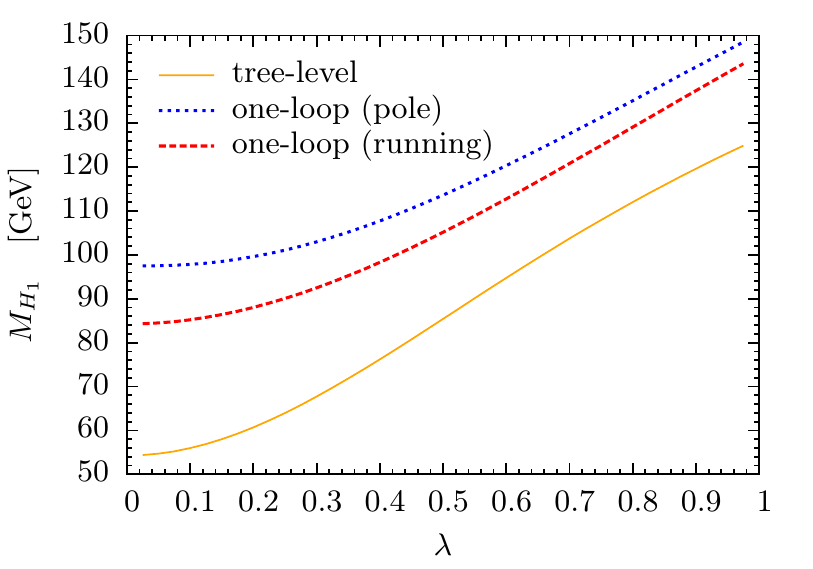}}
\caption{The mass $M_{H_1}$ of the lightest CP-even Higgs boson as
  function of $\lambda$; at tree-level (yellow/full), at one-loop level
  with the top quark pole mass (blue/dotted) and with the running
  $\overline{\mbox{DR}}$ top quark mass
  (red/dashed). From~Ref.~\cite{Ender:2011qh}.} 
\label{fig:oneloopcpeven}
\end{center}
\vspace*{-0.6cm}
\end{figure}

The loop corrections not only considerably alter the Higgs mass values
but they also change the singlet admixtures in the 
various Higgs mass eigenstates and, in the case of CP-violation, their
respective level of CP-violation, with significant phenomenological
implications. For the results shown in the following, the scenarios
have been chosen such that the constraints from LHC Higgs data and
exclusion limits on the SUSY particles are fulfilled. 
The detailed parameter choices and applied constraints can be found in
\cite{Ender:2011qh,Graf:2012hh}. Figure~\ref{fig:oneloopcpeven} 
shows for the CP-conserving case the mass of the lightest CP-even Higgs
boson as function of $\lambda$, at tree-level and at one-loop level
in the $\overline{\mbox{DR}}$ scheme with the top quark mass taken as
the running $\overline{\mbox{DR}}$ mass $m_t=150.6$~GeV at the scale
$Q_0=300$~GeV in one case and
as the pole mass $M_t=173.3$~GeV in the other case. The tree-level
mass increases with rising $\lambda$ due to the 
NMSSM contribution $\sim \lambda^2 \sin^2 2\beta$ from the Higgs
quartic coupling. The one-loop corrections are important, increasing
the $H_1$ mass by up to $\sim 44$~GeV, and strongly depend on the value
of the top quark mass, which reflects the fact that the main part of
the higher-order corrections stem from the top sector. 
An estimate of the missing higher-order corrections can be obtained by
investigating the influence of the renormalization scale $Q_0$. 
The residual theoretical uncertainties due to missing
higher-order corrections can thus be estimated to O(10\%). These
uncertainties as well as the dependence of the corrections on the
value of top quark mass and/or the choice of the renormalization
scheme, of course, get reduced once two-loop corrections are taken into account.

\begin{figure}[b!]
\begin{center}
\includegraphics[width=0.95\linewidth]{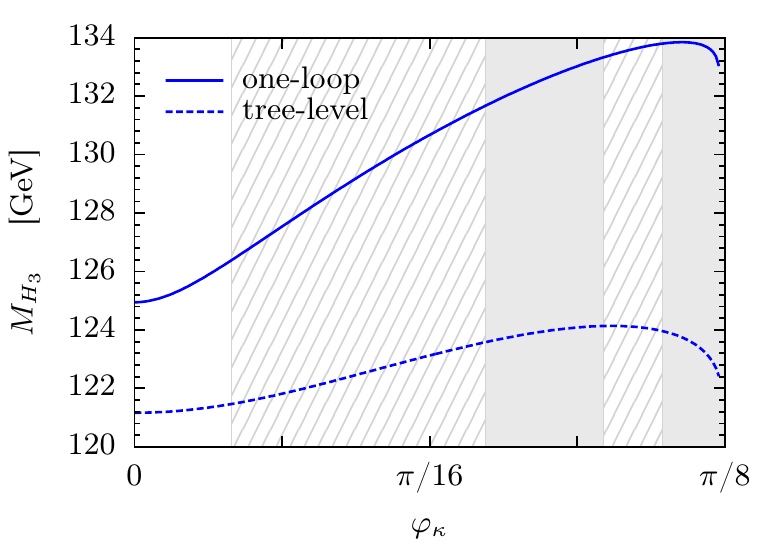} 
\end{center}
\vspace*{-0.5cm}
\caption{Tree-level (dashed) and one-loop (full) mass $M_{H_3}$
 as a function of $\varphi_\kappa$. The exclusion region due to LEP,
 Tevatron and LHC data is shown as grey area, the region with the
 SM-like Higgs boson not being compatible with an excess of data
 around 125 GeV as dashed area. 
Taken from Ref.~\cite{Graf:2012hh}.}
\label{fig:kappavar} 
\end{figure} 
In the NMSSM CP-violation in the Higgs sector can already occur at
tree-level due to a non-vanishing phase $\varphi_y$. 
The effect of a non-vanishing phase $\varphi_\kappa$ is demonstrated in
Fig.~\ref{fig:kappavar} which shows the tree-level and one-loop mass
of $H_3$ which in the chosen scenario corresponds to a
SM-like Higgs boson with mass around 125~GeV. As expected, the mass
exhibits already at tree-level a sensitivity to the CP-violating phase
$\varphi_\kappa$. This dependence is even more pronounced at
one-loop level, changing the $H_3$ mass value by up to 9 GeV for
$\varphi_\kappa \in [0,\pi/8]$. The grey areas are the parameter regions which are
excluded due to the experimental constraints from LEP, Tevatron and
LHC, the dashed region excludes the parameter regions where the
criteria of compatibility with the Higgs excess around 125 GeV cannot
be fulfilled any more. See \cite{Graf:2012hh} for details. The
two-loop corrections at O($\alpha_t \alpha_s$) based on a mixed
$\overline{\mbox{DR}}$-on-shell renormalization scheme 
\cite{Muhlleitner:2014vsa} have been provided for both 
$\overline{\mbox{DR}}$ and on-shell renormalization in the
top/stop sector. For the light Higgs boson masses, the corrections
turn out to be important and are of the order of 5-10\% for the
SM-like Higgs boson, depending on the adopted top/stop renormalization
scheme. An estimate of the remaining theoretical uncertainties due to
missing higher order corrections, by varying the renormalization
scheme in the top/stop sector, shows, that the uncertainty is reduced
from one- to two-loop order. The difference in the mass values of the
SM-like Higgs boson for the two schemes decreases from 15-25\% to
below 1.5\%. These two-loop corrections together with the one-loop
corrections have been implemented in the Fortran package
\textsc{Nmssmcalc} \cite{Baglio:2013iia}, which provides besides the
loop-corrected Higgs boson masses in the real and complex NMSSM also
the NMSSM Higgs branching ratios including the state-of-the-art higher order
corrections and the relevant off-shell decays.

Not only the Higgs boson masses but also the Higgs self-interactions
arise from the Higgs potential, and they can hence not be separated
from each other. In order to consistently describe the Higgs sector
including higher-order corrections, it is therefore not sufficient to
only correct the Higgs boson masses. The trilinear and quartic Higgs
self-interactions have to be evaluated at the same order in
perturbation theory and within the same renormalization scheme as the
Higgs boson masses to allow for a consistent description of the Higgs
boson phenomenology. The trilinear Higgs self-couplings play a role in
the determination of the Higgs boson branching ratios into SM
particles, in the evaluation of Higgs-to-Higgs decays and in Higgs
pair production processes. The one-loop corrections to the trilinear
Higgs self-couplings of the CP-conserving NMSSM in the
Feynman-diagrammatic approach have been evaluated in \cite{Nhung:2013lpa} by
applying the mixed $\overline{\mbox{DR}}$-on-shell renormalization scheme,
used also for the one-loop corrections to the Higgs boson masses
presented here and introduced in \cite{Ender:2011qh}. Table
\ref{tab:dihiggstril} summarizes the Higgs pair production cross
section values for gluon fusion in scenarios where the heavy CP-even
$H_3$ is large enough to allow for the production of a pair of SM-like
Higgs bosons. Here $\sigma_T$ denotes the cross sections calculated
using the effective tree-level trilinear Higgs couplings, while the
cross section values $\sigma_L$ use the effective loop-corrected
trilinear Higgs self-couplings. The differences in the cross sections
due to the inclusion of loop corrections in the trilinear Higgs
self-couplings can be substantial, ranging from nearly 40\% to almost
90\% in terms of the tree-level cross section for the chosen scenarios. 
\begin{table}[h]
 \begin{center}
\begin{tabular}{c|c|c|c}
 \hline
{}
&$\sigma_\text{T} [fb]$ 
&$\sigma_\text{L} [fb]$
&$\delta$\\
\hline \hline
Point 1    & 432.4(3) &  96.08(7) &  -0.78      \\
Point 2    & 181.5(3)  &55.92(2)&      -0.69  \\ 
Point 3   & 533.9(4)   &  265.5(2) & -0.50 \\ 
Point 4   & 413.3(3) &   53.05(4)  & -0.87   \\
Point 5   & 43.24(2)  & 69.05(5)  & 0.60   \\ 
\hline 
\end{tabular}
\caption{\label{tab:dihiggstril}{The total cross sections in fb for $pp\to
     H_iH_i$ through gluon fusion at $\sqrt{s}= 14$~TeV, with $H_i$
     being the SM-like Higgs boson,
     evaluated with tree-level ($\sigma_T$) and loop-corrected
     ($\sigma_L$) effective trilinear Higgs couplings. The
     deviation in the cross sections is quantified by
     $\delta=(\sigma_L-\sigma_T)/\sigma_T$. For details, see
     \cite{Nhung:2013lpa}.}} 
\end{center}
\vspace*{-0.5cm}
\end{table} 

The NMSSM with its increased parameter set in the Higgs sector and in the
soft SUSY breaking Lagrangian allows for a large playground in 
Higgs boson phenomenology. We have performed extensive parameter scans
in the NMSSM by taking into account the constraints from the LHC Higgs
data, from the LHC searches for SUSY particles and the ones arising
from Dark Matter, low-energy observables and from the LEP and Tevatron
data. In Ref.~\cite{King:2012is} we have shown that scenarios can be found that
are compatible with all constraints and can accommodate the
enhanced di-photon final state rate, which was then still present both
in the ATLAS and CMS data. This can be achieved even for rather
light stop masses without large fine-tuning. The scan of Ref.~\cite{King:2012tr}
shows that there is a substantial amount of parameter space with Higgs
boson masses and couplings compatible with the LHC results. If guided
by fine-tuning considerations, the next-to-lightest Higgs boson $H_2$
being the SM-like state is favoured. However, dropping this assumption
also leads to $H_1$ being the SM-like scalar or even scenarios where
$H_1$ and $H_2$ are almost degenerate in mass and close to
125~GeV. The extensive survey of the NMSSM performed in \cite{King:2014xwa}
revealed that the natural NMSSM, which we defined to be characterized
by rather small $\kappa$ values, low $\tan\beta$ and an overall Higgs
spectrum below about 530~GeV, can be tested at the next run of the LHC with a
c.m.~energy of 13~TeV. This relies on exploiting also Higgs production
from Higgs-to-Higgs decays and Higgs decays into a Higgs and gauge
boson pair. Focusing subsequently on these cascade decays, we found
that within the NMSSM exotic final state signatures with 
multi-photon and/or multi-fermion final states at significant rates
can be possible arising from one or several Higgs cascade decays. They
furthermore give access to the trilinear Higgs self-couplings. These
interesting and sometimes unique signatures should be taken into
account when designing analysis strategies for the LHC. 

\section{Higgs Coupling Measurements and Implications for New Physics Scales}
\label{sec:couplings}
According to the present  data the observed Higgs particle is in good
agreement with SM expectations. The experimental uncertainties are
still large though and allow for interpretations in models beyond the
SM as we have seen in the previous sections. The high-energy and
high-luminosity run of the LHC will increase the precision on the
data. Thus the precision on the Higgs couplings will improve from at present
several tens of percent to about 10\% in the high-luminosity (HL-LHC)
option \cite{Lafaye:2009vr,Klute:2012pu,Plehn:2012iz,Bechtle:2014ewa,Klute:2013cx}. A future $e^+e^-$
linear collider (LC) \cite{Bechtle:2014ewa,Klute:2013cx,AguilarSaavedra:2001rg,Accomando:2004sz,Djouadi:2007ik,Baer:2013cma} can
improve the accuracy to about 1\%, {\it 
  cf.}~Table~\ref{tab:coupprec}. Deviations in the interactions of the
Higgs boson from their SM values can arise if the Higgs mixes with
other scalars, if it is a composite particle or a mixture between an
elementary and composite state (partial compositeness) or through loop
contributions from other new particles. Depending on the strength and
type of the coupling between the new physics and the Higgs boson, the
limits derived from the Higgs measurements can exceed those from
direct searches, EW precision measurements or flavor physics. Higgs
precision analyses can thus be sensitive to new physics residing at
scales much higher than the VEV and open a unique window to new physics sectors
that are not yet strongly constrained by existing results. The
effective field theory approach allows to study a large class of BSM models in
terms of a well defined quantum field theory. It cannot account,
however, for effects that arise from light particles or from Higgs
decays into new non-SM particles. In order to give a complete picture
of BSM effects in the Higgs sectors, therefore also specific BSM
models capturing such features have to be studied. In the following
a few examples for both approaches shall be highlighted, based on
Ref.~\cite{Englert:2014uua}, where more details and further investigated models
can be found.  
\begin{table}[t!]
\begin{center}
\begin{tabular}{l||c|c||c|c||c}
\hline
{\small coupl.}	    &  {\small LHC}  & {\small HL-LHC}	       & {\small LC}
& {\small HL-LC} &  {\small comb.}\\
\hline\hline
$hWW$		    &  0.09 & 0.08	       & 0.011	   & 0.006 & 0.005	    \\
$hZZ$		    &  0.11 & 0.08	       & 0.008	   & 0.005 & 0.004	    \\
$htt$		    &  0.15 & 0.12	       & 0.040	   & 0.017 & 0.015	    \\
$hbb$		    &  0.20 & 0.16	       & 0.023	   & 0.012 & 0.011	    \\
$h\tau\tau$	    &  0.11 & 0.09	       & 0.033	   & 0.017 & 0.015	    \\
\hline
$h\gamma\gamma$     &  0.20 & 0.15	       & 0.083	   & 0.035 & 0.024	    \\
$hgg$		    &  0.30 & 0.08	       & 0.054	   & 0.028 & 0.024	    \\
\hline
$h_{\scriptsize \mathrm{invis}}$   &	---  & ---		& 0.008     & 0.004 & 0.004	     \\
\hline
\end{tabular}
\vspace*{0.2cm}
\caption{Expected accuracy at the 68\% C.L.\ with which fundamental and derived
  Higgs couplings can be measured; the deviations are defined as
  $g=g_{\scriptsize \mbox{SM}} [1\pm\Delta]$ compared to the Standard Model at the LHC/HL-LHC (luminosities
  300 and 3000 fb$^{-1}$), LC/HL-LC (energies 250+500~GeV / 250+500~GeV+1~TeV and luminosities
  250+500 fb$^{-1}$ / 1150+1600+2500 fb$^{-1}$), and in combined analyses of HL-LHC and HL-LC.
  For invisible Higgs decays the upper limit on the underlying
  couplings is given. Taken from \cite{Englert:2014uua}.}
\label{tab:coupprec}
\end{center}
\vspace*{-0.6cm}
\end{table}

Based on operator expansions
\cite{Burges:1983zg,Leung:1984ni,Buchmuller:1985jz,Grzadkowski:2010es} deviations from the SM 
coupling values can be estimated to be of the order of
\begin{equation}
g = g_{\scriptsize \mbox{SM}} [1 + \Delta] \; : \; \Delta = {\cal O}
(v^2/\Lambda^2) \;,
\end{equation}
where $v$ denotes the VEV of the SM field and $\Lambda \gg v$ the
characteristic BSM scale. Note that this does not hold in case the
underlying model violates the decoupling theorem. Assuming
experimental accuracies of $\Delta = 0.2$ down to $0.01$ implies a
sensitivity to scales of order $\Lambda \sim 550$~GeV up to
2.5~TeV. The smaller bound is complementary to direct LHC searches
whereas the larger of the two bounds in general exceeds the direct LHC
search range. If the Higgs coupling deviations are due to vertex
corrections generated by virtual contributions of new particles, a
suppression factor of $1/(16\pi^2)$ has to be taken into account in
addition to potentially small couplings between the SM and the new
fields. Hence only new scales not in excess of about $M< v/(4\pi
\sqrt{\Delta})\sim 200$~GeV can be probed, which in most models is much
less than the direct search reach at the LHC. 

\begin{figure}[h]
\begin{center}
\includegraphics[width=0.5\textwidth]{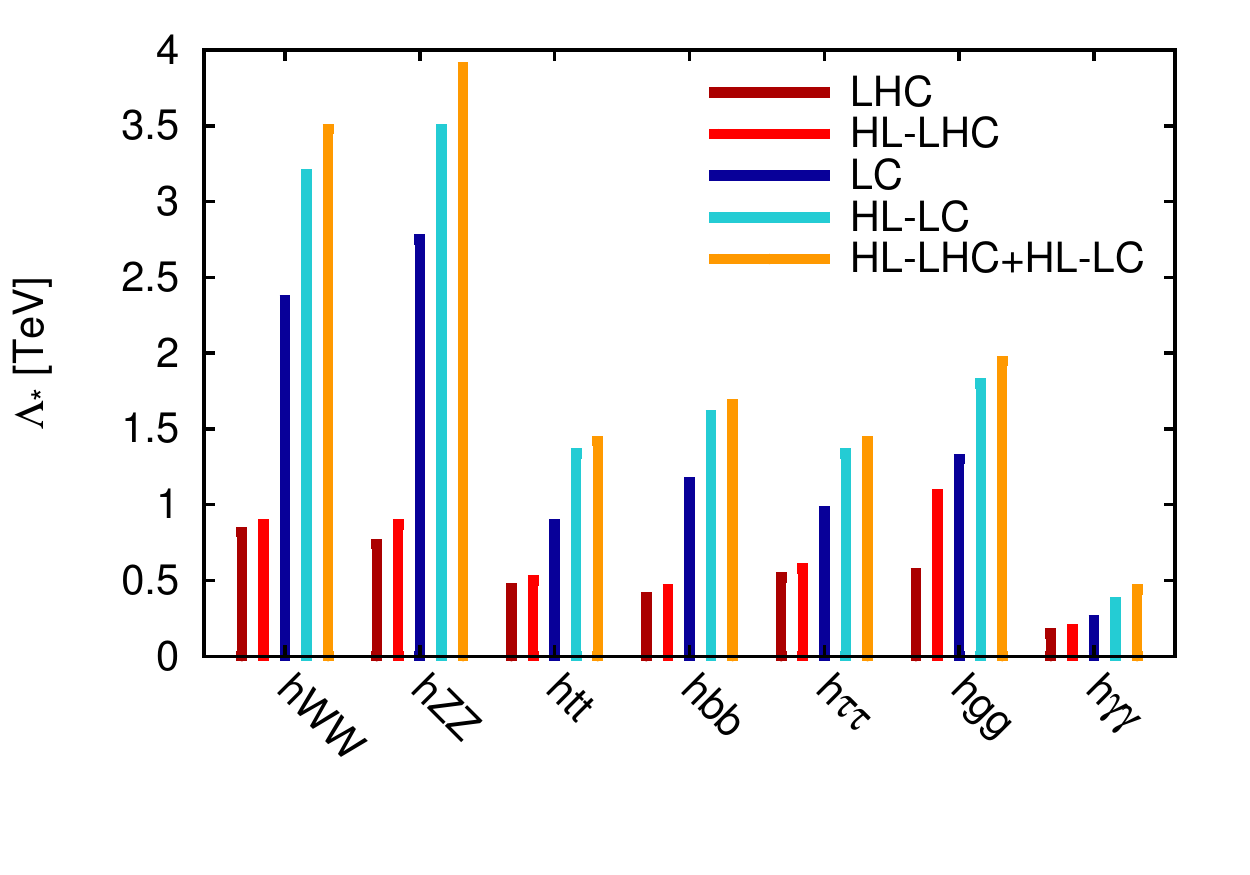}
\vspace{-1.2cm}
\caption{Effective NP scales $\Lambda_\ast$ extracted from the
	 Higgs coupling measurements collected in
	 Table~\ref{tab:coupprec}. (The ordering of
	 the columns from left to right corresponds to the legend
	 from up to down.) For details, see \cite{Englert:2014uua}.}
\label{fig:limscales}
\end{center}
\vspace*{-0.6cm}
\end{figure}

The extracted limits on the effective scales $\Lambda_*$ from the
contributions of the dimension-6 operators taking into 
account the coupling precisions of Table~\ref{tab:coupprec} are shown in
Fig.~\ref{fig:limscales}. They have been obtained with \textsc{Sfitter}
\cite{Lafaye:2009vr,Klute:2012pu,Plehn:2012iz} after defining the
effective scales $\Lambda_*$ that are 
obtained by factoring out from the operators typical coefficients like
couplings and loop factors. Furthermore, in the loop-induced couplings
to the gluons and photons only the contributions from the contact
terms are kept. The effects of the loop terms are already disentangled
at the level of the input values $\Delta$. 

\begin{figure*}
\makebox[0.23\textwidth]{\centering type-I}
\makebox[0.23\textwidth]{\centering type-II}
\makebox[0.23\textwidth]{\centering lepton-specific}
\makebox[0.23\textwidth]{\centering flipped}\\[-1em]
\mbox{
\raisebox{-\height}{\includegraphics[width=0.23\textwidth]{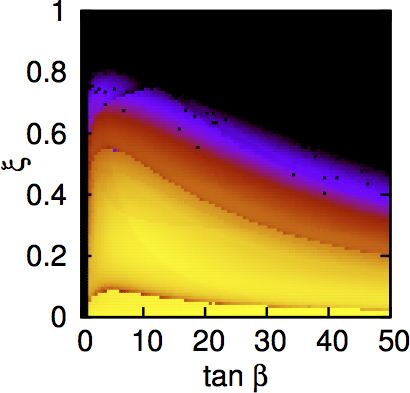}}
\raisebox{-\height}{\includegraphics[width=0.23\textwidth]{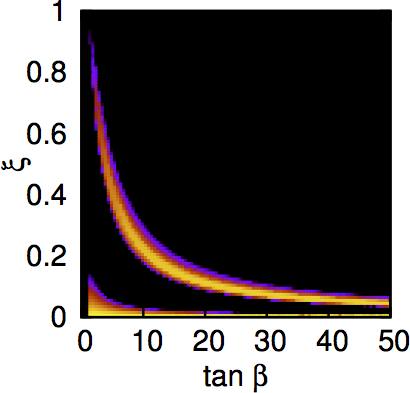}}
\raisebox{-\height}{\includegraphics[width=0.23\textwidth]{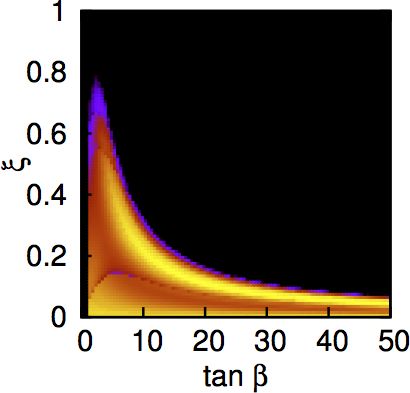}}
\raisebox{-\height}{\includegraphics[width=0.23\textwidth]{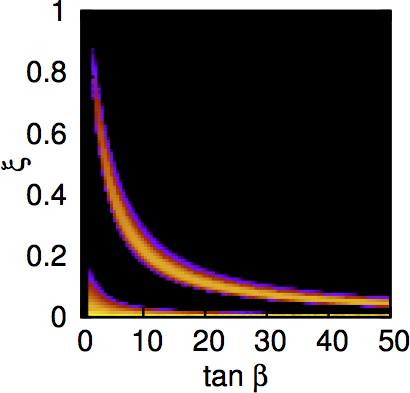}}
\raisebox{-\height}{\includegraphics[width=0.035\textwidth]{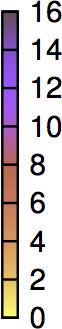}}}
\caption{Allowed ranges for the modification parameter $\xi$ in a 2HDM in the
decoupling limit, based on data from Ref.~\cite{ATLASCMS2}. The plots show the
correlated relative log-likelihood $-2\,\Delta(\log\cal L)$ as a function of
$\tan\beta$, from Ref.~\cite{Lopez-Val:2013yba}.}
\label{fig:2HDM}
\end{figure*}
As a last example, the interpretation of the current Higgs coupling
measurements in terms of a 2-Higgs-Doublet Model (2HDM) is
shown. 
Here the Higgs coupling modifications are due to mixing effects, with
the physical states being mixtures of the components of two doublets
$\phi_1$ and $\phi_2$ \cite{Lee:1973iz,Flores:1982pr,hhguide,Branco:2011iw,Gunion:2002zf}. 
The scalar potential can be cast into the form
\begin{eqnarray}
V &=& m_{11} |\phi_1|^2 + m_{22}^2 |\phi_2|^2 - m_{12}^2(\phi_1^\dagger \phi_2 +
\text{h.c}) \nonumber \\
&+& \lambda_1 |\phi_1|^4 + \lambda_2 |\phi_2|^4 
+ \lambda_3 |\phi_1|^2 |\phi_2|^2 + \lambda_4 |\phi_1^\dagger
\phi_2|^2 \nonumber \\
&+& \frac{1}{2}\lambda_5 [(\phi_1^\dagger \phi_2)^2 + \text{h.c}]\,. 
\label{eq:2hdm_pot}
\end{eqnarray}
By imposing a global $\mathbb{Z}_2$ discrete symmetry, under which
$\phi_{1,2} \to \mp \phi_{1,2}$, it can be achieved that one type of
fermions couples only to one Higgs doublet. This ensures the natural
suppression of flavor-changing neutral currents. In the potential
Eq.~\eqref{eq:2hdm_pot} such a symmetry has been assumed, softly broken 
by the term $\propto m_{12}^2$. The Higgs fields
acquire vacuum expectations values,
$v_1$ and $v_2$, with $v^2_1+v^2_2=v^2$ and $\tan\beta = v_2/v_1$.
Depending on the $\mathbb{Z}_2$ charge assignments, the following four cases of
coupling the Higgs doublets to fermions are possible~\cite{Barger:1989fj}:
\begin{itemize}
\item \underline{type I:} all fermions couple only to $\phi_2$;
\item \underline{type II:} up-/down-type fermions couple to
$\phi_2$/$\phi_1$, respectively;
\item \underline{lepton-specific:} quarks couple to $\phi_2$ and charged
leptons couple to $\phi_1$;
\item \underline{flipped:} up-type quarks and leptons couple to $\phi_2$ and
down-type quarks couple to $\phi_1$.
\end{itemize}

Note that the MSSM is a special case of the general 2HDM type-II.
After EWSB the Higgs sector features five physical Higgs states, two
neutral CP-even ones $h^0, H^0$, one neutral CP-odd one $A^0$ and two
charged Higgs bosons $H^\pm$.  
In the so-called decoupling limit the masses of the heavy Higgs bosons
$H^0, A^0$ and $H^\pm$ are much larger than $v$ and the physics of the
light Higgs boson $h^0$ can be described by an effective theory
\cite{Gunion:2002zf}. 
The properties of the lightest CP-even Higgs boson $h^0$ are
close to the SM. Figure~\ref{fig:2HDM} shows the shape of
the decoupling limit in the different model setups. Type-I models are
preferred in a comparably wide parameter range which is due to the
fact that it separates Higgs couplings to gauge bosons and
fermions. This makes it easy to accommodate
the slightly enhanced $H \to \tau \tau$ rate. So far none of the
analyses based on the Higgs couplings measured at the LHC has shown a
clear sign for such mixing effects. Further discussions within
typical scenarios and models that are archetypal examples for a much
larger class of models can be found in \cite{Englert:2014uua}. 

\section{Conclusions}
\label{sec:conclusions}
While the properties of the new particle recently discovered at the LHC  are
consistent with those expected for the Higgs boson of the
Standard Model, the present experimental precision still leaves
room for interpretations within BSM extensions. These can be either 
strongly interacting theories, or rely on weak interactions as \textit{e.g.}~supersymmetric models. In order to test a wide range of new physics
scenarios in a more model-independent way, the effective Lagrangian approach can be applied for the 
interpretation of the data. Such an approach is valid as long as the new
physics scale is well above the Higgs mass value. The interpretation of LHC data through effective 
Lagrangians must
be complemented by analyses within specific models in order to take
into account effects from low-lying resonances. 

Composite Higgs
models are specific realizations of models based on strong
dynamics. They are heavily constrained by electroweak precision data, however still in
accordance with the LHC data, in particular when new heavy quarks
are taken into account. We have discussed in detail the compatibility of
composite Higgs models with the LHC results and the role
of new heavy quarks in the loop-induced single and double Higgs
production processes through gluon fusion. 

Supersymmetric models, on the other hand, are
weakly interacting. In the minimal version, the MSSM, a sufficiently
large Higgs mass of 125 GeV can only be achieved for heavy stops and/or
large mixing, whereas the next-to-minimal model (NMSSM) with its enlarged parameter space can
accommodate the Higgs mass more easily. In order to distinguish various SUSY
models from each other, and from other BSM extensions, higher-order
corrections are essential. Only the high-precision calculations 
allow to correctly interpret the experimental data. We have reported on recent progress
in higher-order calculations to MSSM Higgs boson production in
association with heavy quarks and on the calculation of NMSSM Higgs boson masses and
self-interactions. 

Finally, we have presented the conclusions that can be drawn on the scale of new
physics from high-precision measurements of the
Higgs couplings at the upcoming LHC runs, and at a future $e^+e^-$ linear collider.

\section*{Acknowledgements}
We would like to thank our collaborators 
J.\ Baglio, A.\ Biek\"otter,  S.\ Dittmaier, A.\ Djouadi, R.\ Contino, C.\ Englert, J.R.\ Espinosa, M.\ Flechl, A.\ Freitas, M.\ Ghezzi, M.\ Gillioz, M.\ Gomez-Bock, T.\ Graf, R.\ Gr\"ober, C.\ Grojean, P.\ H\"afliger, R.\ Harlander, A.\ Kapuvari, A.\ Knochel, W.\ Kilian, S.F.\,King, R.\,Klees, D.\,Liu, M.\,Mondragon, A.\ M\"uck, R.\ Nevzorov, D.T.\ Nhung, R.\ Noriega-Papaqui, I.\ Pedraza, T.\ Plehn, J.\ Quevillon, M.\ Rauch, 
F.\ Riva, H.\ Rzehak, E.\ Salvioni, T.\ Schl\"uter, M.\ Schumacher, M.\ Spira, J.\ Streicher, M.\ Trott, M.\ Ubiali, M.\ Walser, K.\ Walz, and P.M.\ Zerwas.

This work was supported by the Deutsche Forschungsgemeinschaft through the collaborative research centre SFB-TR9 ``Computational Particle Physics", 
and by the U.S. Department of Energy under contract DE-AC02-76SF00515. 
MK is grateful to SLAC and Stanford University for their hospitality. 




\nocite{*}



\end{document}